\documentclass[review]{elsarticle}
\usepackage{amssymb,lineno}
\usepackage[a4paper,margin=0.5in]{geometry}
\usepackage{color}
\usepackage{graphics}
\usepackage{mathrsfs}
\usepackage{amsmath,bm}
\usepackage{subfigure}
\usepackage{booktabs}
\usepackage{amsthm}
\usepackage{enumerate}
\usepackage{listings}
\usepackage{setspace}
\usepackage{appendix}
\usepackage{hyperref}
\usepackage{bm}
\usepackage{booktabs}
\graphicspath{{TikzFig/}{picture/}}
\begin{document}

\title{Discontinuity Computing Using Physics-Informed Neural Network}

\author{Li Liu$^1$,Shengping Liu$^1$, Hui Xie$^1$, Fansheng Xiong$^{1}$,Tengchao Yu$^1$, Mengjuan Xiao$^1$, Lufeng Liu$^1$, Heng Yong$^{1,*}$}

\address{$^1$ Institute of Applied Physics and Computational Mathematics, Beijing 100094, China}

\begin{abstract}
  Simulating discontinuities is a long standing problem especially for shock waves with strong nonlinear feather. Despite being a promising method, the recently developed physics-informed neural network (PINN) is still weak for calculating discontinuities compared with traditional shock-capturing methods. In this paper, we intend to improve the shock-capturing ability of the PINN. The primary strategy of this work is to weaken the expression of the network near discontinuities by adding a gradient-weight into the governing equations locally at each residual point. This strategy allows the network to focus on training smooth parts of the solutions. Then, automatically affected by the compressible property near shock waves, a sharp discontinuity appears with wrong inside shock transition-points compressed into well-trained smooth regions as passive particles. We study the solutions of one-dimensional Burgers equation and one- and two-dimensional Euler equations. Compared with the traditional high-order WENO-Z method in numerical examples, the proposed method can substantially improve discontinuity computing.


\end{abstract}

\begin{keyword}
  PINN, Shock capturing, Compressible flow, Euler equations, Discontinuity  calculation
\end{keyword}

\maketitle

\section{Introduction}

Calculating shock waves and other discontinuities sharply and without oscillations is essential for solving hyperbolic equations. The first study on shock-capturing dates back to the development by von Neumann and Richtmeyer \cite{vonneumann1950method}, who introduced artificial viscosity into a staggered Lagrangian scheme to solve compressible flows. Nowadays, various advanced and high-order methods allow to simulate problems with shock waves. These methods include essential non-oscillatory (ENO) \cite{harten1987uniformly}, weighted ENO (WENO) \cite{jiang1996efficient}, and discontinuous Galerkin \cite{cockburn1998local} methods. More information about the development of shock-capturing methods can be found in \cite{pirozzoli2011numerical,zhang2015review}.

Owing to their rapid development, machine learning and neural networks (NNs) have been used to solve partial differential equations \cite{raissi2019physics,pang2019neural,lye2020deep,magiera2020constraint,huang2022neural}.  
In particular, the physics-informed neural network (PINN) attracts much research attention. PINN encodes partial differential equations (PDEs) or other model equations as one of its components. Given its generality, PINN has been used for solving equations in many fields \cite{cuomo2022scientific}. For hyperbolic equations, Patel et al. \cite{patel2022thermodynamically} proposed a PINN that can discover thermodynamically consistent equations ensuring hyperbolicity for inverse problems in shock hydrodynamics. Mao et al. \cite{mao2020physics} studied one- (1D) and two-dimensional (2D) Euler equations with shock waves and used clustered training samples around a high gradient area to improve the solution accuracy in that area while preventing error propagation to the entire domain. 
Jagtap et al. \cite{jagtap2020conservative} proposed the conservative PINN that splits the computing domain into several small subdomains using different NNs to solve Burgers and Euler equations. Jagtap et al. \cite{jagtap2022physics} studied inverse problems in supersonic flows.  
The above mentioned studies have shown the effectiveness of PINN to handle inverse problems with prior information about the development of the flow structures, such as density gradients. However, to study forward problems, the applicability of the original PINN has been limited to simple problems such as tracking moving shock waves. Based on traditional methods,  Patel et al. \cite{patel2022thermodynamically} constructed a mesh-based control-volume PINN and introduced entropy and total-variation-diminishing conditions into the network. Papados \cite{papadossolving} simulated the shock-tube problem with the computing domain extended, obtaining outstanding results without introducing non-physics viscosity terms into the equations.

We aim to enhance the shock-capturing ability of a PINN, especially for complex problems with shock generation. However, the shock wave has zero thickness in compressible inviscid flow theoretically. Thus, it cannot be governed by the strong form of the differential equations given its infinite gradient but instead controlled by a physical process from the left and right regions (Rankine-Hugoniot conditions). Besides, there are also no theory that can guarantee NNs approximate any first-order discontinuous functions.  Hence, if residual points wrongly fall inside a shock region, large equation losses exist in those points given their large gradients. These points are called {\bf{transition-points}} \cite{2014Multistep} in this paper. A NN may focus on handling transition-points as they carry most of the loss. However, a NN cannot increase the gradient to decrease the thickness of the shock owing to the reason talked above. On the other hand, the physics process compresses the region over time. As a result,  transition-points fall into a paradoxical status, in which increasing the gradient seriously increases the total loss because the points are not governed by the equations. However, decreasing the gradient also increases the total loss as it conflicts with physical compression and away from the real solution. More seriously, as the total loss is the sum or average of every point, the transition points attract training and influence the convergence of other points in smooth regions. Here, we can compare the above process with a traditional high-order method, such as the finite-volume WENO method, to illustrate the difficulty of PINN. In a WENO method, when a cell is inside a discontinuous region, the order of the scheme automatically reduces to no more than the second-order. Thus, a large dissipation is appended into the cell to obtain a non-oscillatory result. However, the given cell with large dissipation and consequently large error does not influence the accuracy order of other cells beyond the discontinuity.

We introduce a `retreat to advance' strategy into PINN. To break the paradoxical status of transition-points and obtain a sharp discontinuity, PINN avoids training the shock waves and focuses on training other smooth regions by weakening the network expression in strong compression regions. Then, the compressing property near shock waves allows a sharp and exact shock to appear. This is done by multiplying a local positive and compression-related weight to the governing equations to adjust the expressions of the NN in different residual points. 

The remainder of this paper is organized as follows. In Section 2, we detail the weighted-equation (WE) method. Then, various 1D and 2D forward examples are studied to show the effectiveness of the proposed method. Finally, conclusions are drawn in Section 4.

\section{Method}
\subsection{PINN for conservative hyperbolic PDE}
We consider the following conservative hyperbolic PDE
\begin{equation}
  \frac{\partial {\mathbf U}({\mathbf x})}{\partial t} + \nabla \cdot {\mathbf F}({\mathbf U}) = 0, \quad {\mathbf x}=(t,x_1,x_2,\cdots)\in \Omega,\label{eq:pde}
\end{equation}
with the initial and boundary conditions (IBs)
\begin{equation}
  {\rm IB}({\mathbf U},{\mathbf x}) = 0 \quad  \text{on} \quad \partial \Omega,
\end{equation}
and we treat the initial condition in the same way with Dirichlet boundary condition.

PINN mainly consists of two parts. The first part is a NN $\hat{\mathbf U}({\mathbf x};{\bm \theta}) $  to approximate the relation of ${\mathbf U}({\mathbf x})$ with trainable parameters ${\bm \theta}$. The second part is informed with the governing equations and the IB conditions to train the NN. The calculation of  ${\partial}/{\partial t}$ and $\nabla \cdot $ in the PDE is performed by automatic derivative evaluation. More details about the PINN for concection PDE are available in\cite{mao2020physics,papadossolving}.
\begin{figure}
  \centering
\includegraphics[width = 12cm]{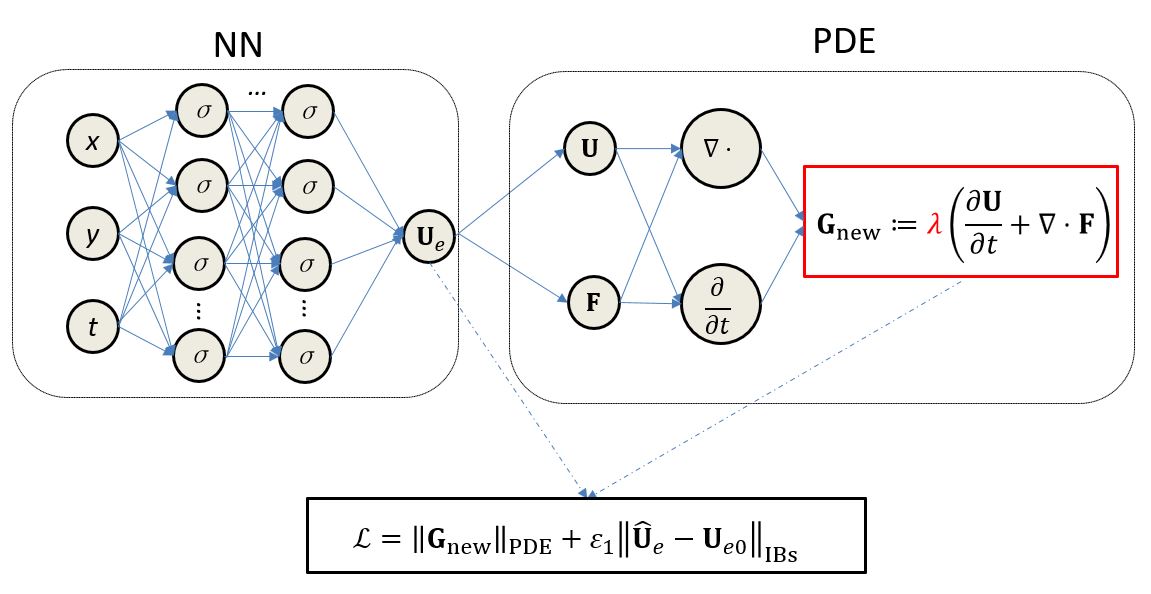}
\caption{Architecture of PINN-WE for solving conservative hyperbolic equations. }\label{fig:PINNWE}
\end{figure}

The loss function used to train $\hat{\mathbf U}({\mathbf x};{\bm \theta})$ contains at least two parts to define the problem.
 One is controlled by the equations, and the other one is given by the
IBs of the problem,
\begin{equation}\label{eq:4}
  \mathcal{L} = \mathcal{L}_{\rm PDE} + \varepsilon_1 \mathcal{L}_{\rm IBs}.
\end{equation}
To define the loss, we choose a set of residual points inside the domain $\Omega$ and another set of points in $\partial \Omega$ as $\mathcal{S}_{\rm PDE}$ and $\mathcal{S}_{\rm IBs}$, respectively. 
Then 
\begin{equation}
  \mathcal{L} = \frac{1}{|\mathcal{S}_{\rm PDE}|}\sum_{{\mathbf x}_i \in \mathcal{S}_{\rm PDE}}{\bf G}_i^2 + \varepsilon_1\frac{1}{ |\mathcal{S}_{\rm IBs}|} \sum_{{\mathbf x}_i \in \mathcal{S}_{\rm IBs}} (\hat{\bf U}_{e,i} - {\bf U}_{e0,i})^2,
\end{equation}
where $ {\bf G}_i: = \partial_t \hat{\bf U}({\bf x}_i) + \nabla \cdot {\bf F(\hat{U}(x_i))}$ and ${\bf G}_i=0$ is the governing equations at residual point ${\bf x}_i \in \mathcal{S}_{\rm PDE} $ and ${\bf U}_{e0,i}$ represents the given initial IB at residual point   ${\bf x}_i \in \mathcal{S}_{\rm IBs} $.

$\varepsilon_1$ is the weight to adjust the  confinement strength  of IBs\cite{jagtap2022physics,yu2022gradient,papadossolving}. A common condition is giving more weight to the points on boundaries.

In each term of a loss function, averaging the residual across all residual points is common to obtain the total training loss. Averaging may be suitable and convenient for problems with smooth solutions. However, when a discontinuity occurs, the gradient becomes theoretically infinity and cannot be described directly by differential equations. Consequently, transition-points defined inside the discontinuity may introduce large errors and function loss. We will show this in the following case.

\subsection{Analysis of transition-points based on Burgers equation}
Here we consider inviscid Burgers equation which is a scalar nonlinear hyperbolic equation to illustrate the training problem of transition-points. 

\begin{figure}
  \centering
  \subfigure[Loss history with PINN]{
\includegraphics[width = 9cm]{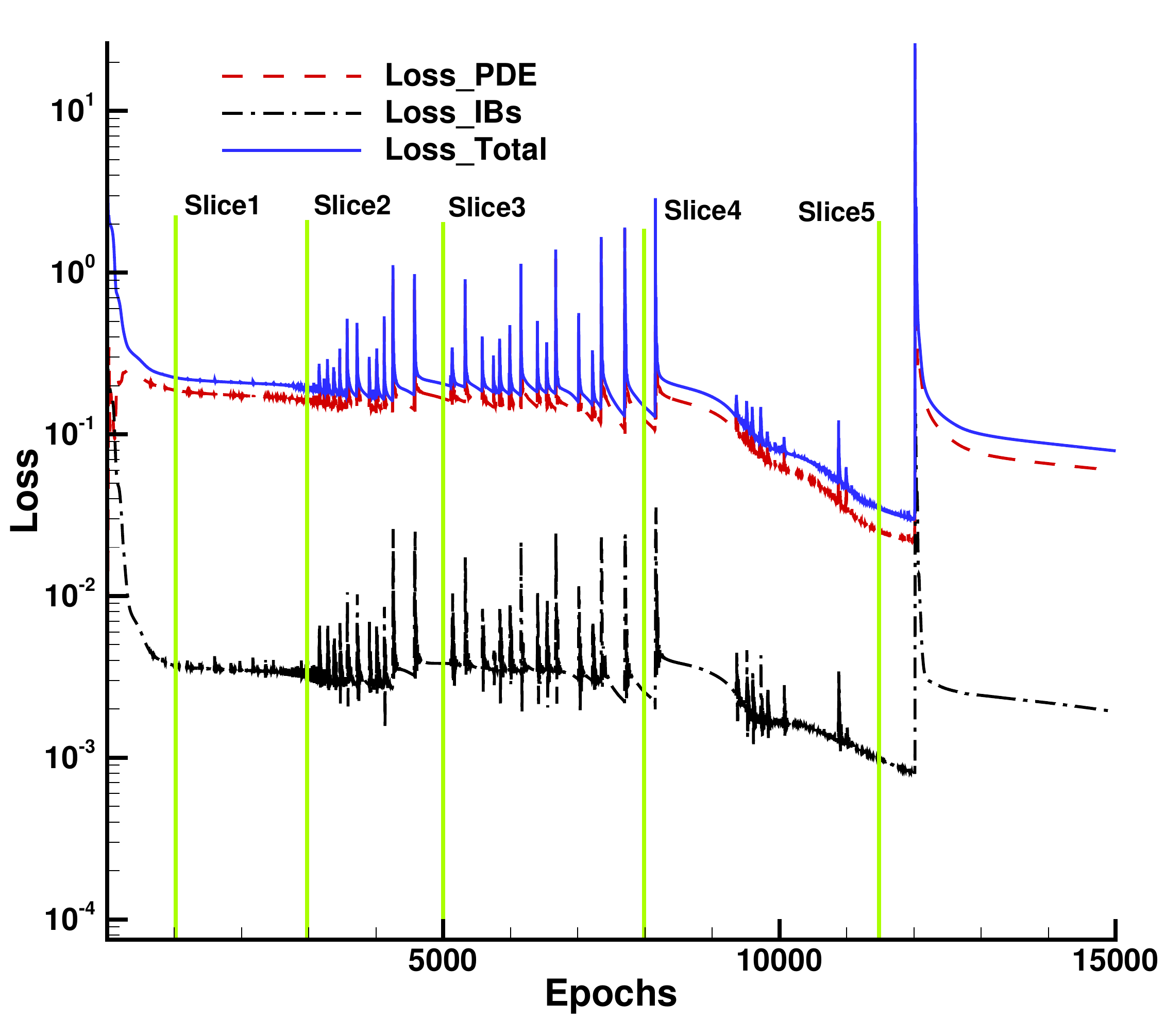}}
  \subfigure[$u$ at $t=1$ at different epochs]{
\includegraphics[width = 9cm]{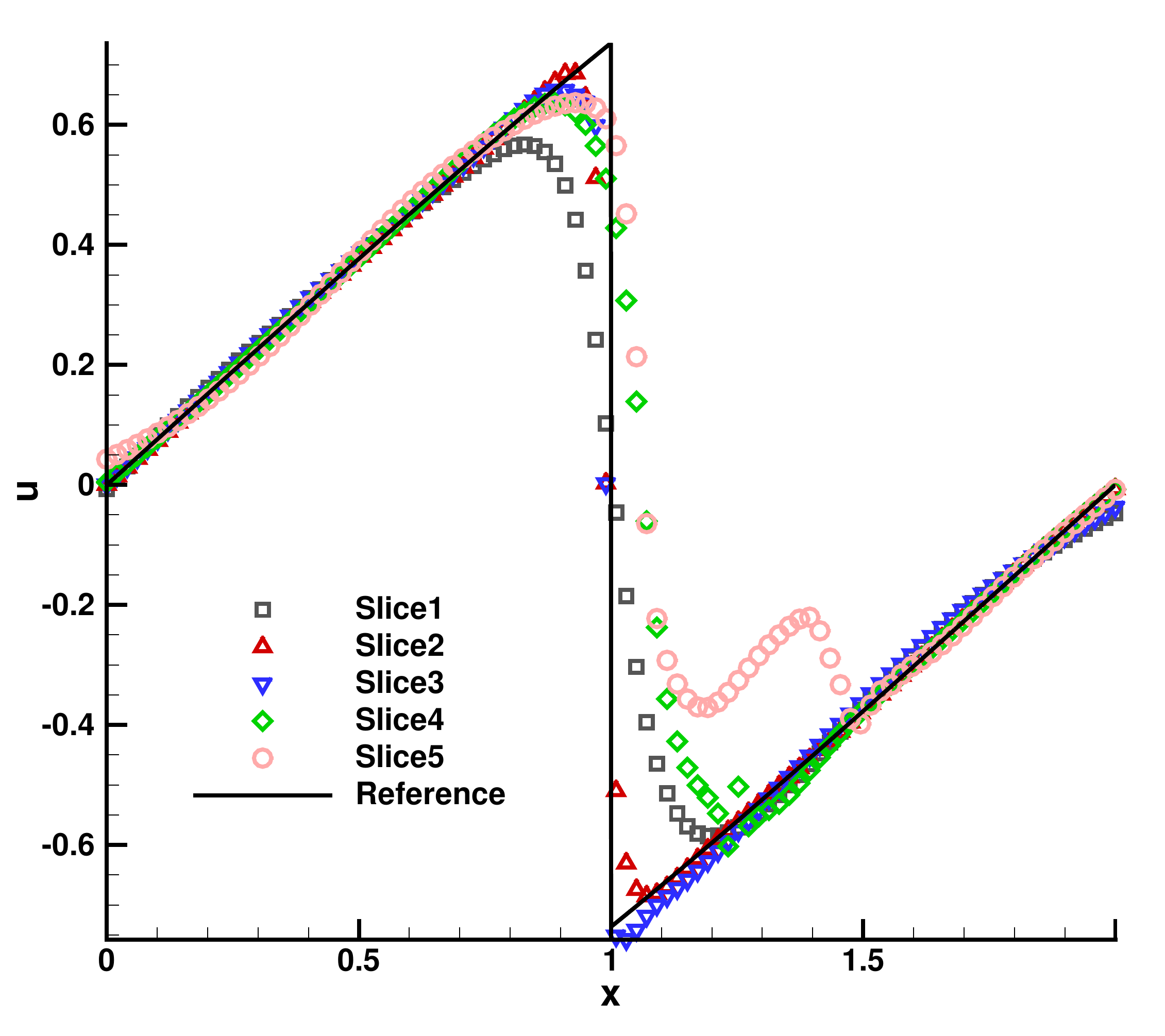}}
  \subfigure[Residual of $x$ at $t=1$ at different epochs]{
\includegraphics[width = 12cm]{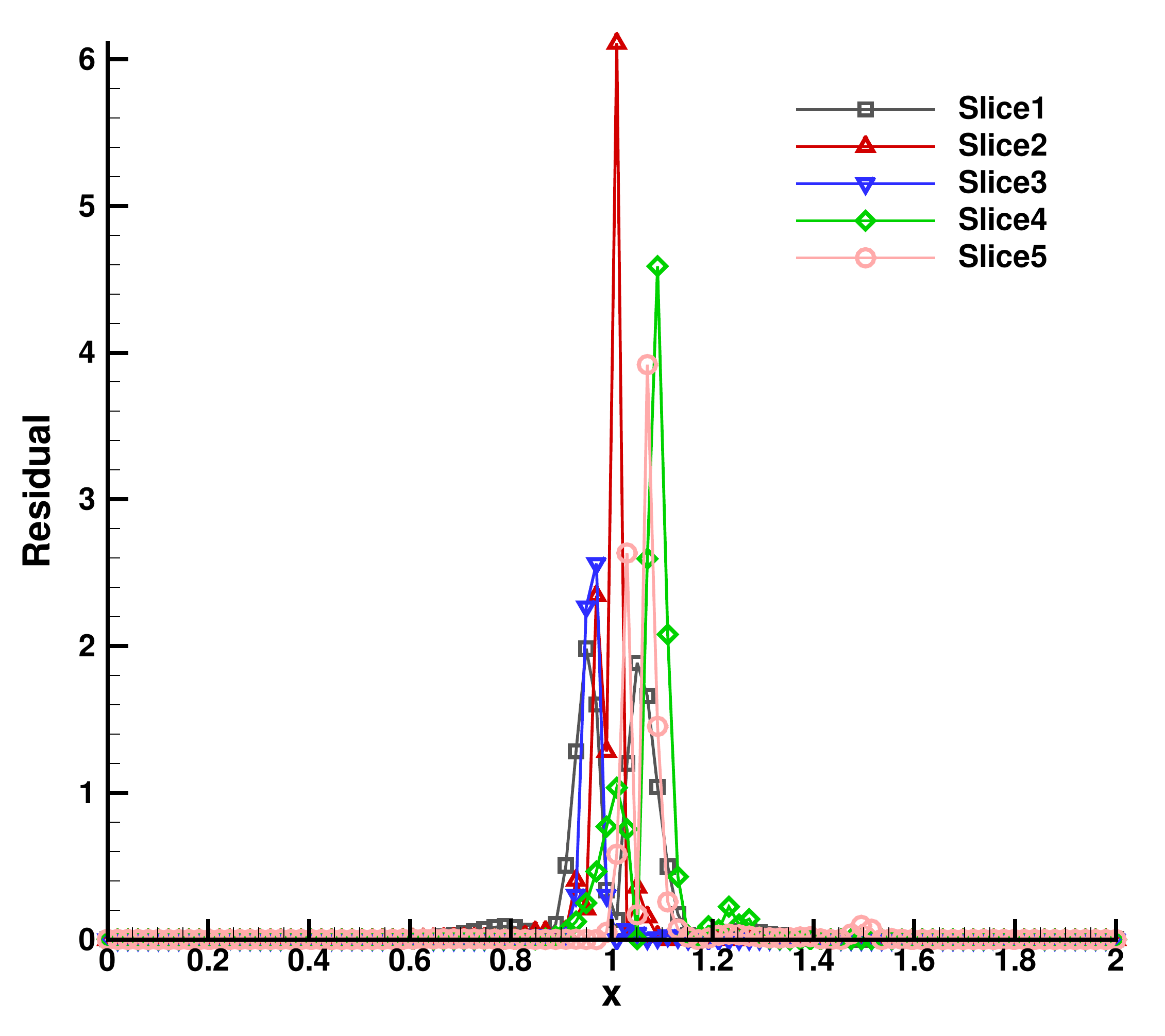}}
\caption{Results of Burgers equation with PINN. NN with 4 hidden layers and 30 neurons per layer. The PDE residual points are set with uniform $100\times 100$ grids on $X \times T$ space and the IBs points are set with uniform $100$ grids.The optimizer is Adam with a learn rate $0.001$. The reference result is given with WENO-Z on a refine mesh with 10000 spatial grids.}\label{fig:burgers_1}
\end{figure}

The equation and IBs are given as 
\begin{equation}
  \begin{aligned}
 & \frac{\partial u}{\partial t} + \frac{\partial (u^2/2)}{\partial x} = 0, \quad x\in [0,2], \quad t \in [0,1],\\
 & u(0,x) = -{\rm sin}(\pi (x-1)),\\
 & u(t,0) = u(t,2) = 0.
  \end{aligned}\label{eq:burgers}
\end{equation}
We first solve this problem with a traditional PINN introduced in above. Fig.\ref{fig:burgers_1} gives the loss history and $u$ and the residual as functions of $x$ at $t=1$ at different training epochs (1000,3000,5000,8000 and 11500). It is clearly that PINN first tends to reach a global smooth solution to fit the IBs to get a small $\mathcal{L}_{\rm IBs}$ (Slice 1). And then the training attends to further deduce the residual to reach the exact solution of the problem. However, when the transition-points reach a relatively high gradient (Slice 2), they nearly carry all the function loss $\mathcal{L}_{\rm PDE}$ (Fig,\ref{fig:burgers_1}.c). Then the training will fall into a paradoxical status:  no matter increase (Slice 3) or decrease the gradient (Slice 4) of transition-points will increase the function loss. This is due to those points can not be controlled by the equation directly, then decrease the gradient will remove from the exact solution but increase the gradient also will increase the residual. As shown in the loss history, after 3000 epochs, the training struggles in those transition-points and can not decrease the total loss effectively. More seriously, as a global method, the total loss also decide the error size of other regions. 


\subsection{PINN-WE}

Based on above analysis of Burgers equation, the training will fall into straggling due to the transition-points if we take an average of each point residual into the total loss.  We propose an weighted equation method to weaken the effect of points in highly compressible regions by assigning a local positive weight $\lambda$ to the governing equations. 
This method is based on the fact that strong discontinuity is formed by the convergence of the characteristic lines.  As we weaken the expression of transition-points, the NN will focus on training the smooth regions and get a high accuracy in those region, then automatically effected by the compression property of the strong discontinuity, the transition-points will be compressed into smooth region. Then a sharp and exact discontinuity solution will appear with the training.

For a general conservative equation (Eq. 1), 
 \begin{equation}\label{eq:5}
  {\bf G}_{\rm new}: = \lambda(\frac{\partial {\bf U}}{\partial t}+ \nabla\cdot {\bf F}),
 \end{equation}
 and ${\bf G}_{\rm new} = 0$ is WE. Then, 
${\bf G}_{\rm new} = 0$ has the same solutions as $G= 0$ if $\lambda$ is always positive. In addition, we can adjust the NN expression in different points by the design of gradient-dependent weight $\lambda$. Correspondingly, the  new loss is defined as
\begin{equation}
  \mathcal{L} = \frac{1}{|\mathcal{S}_{\rm PDE}|}\sum_{{\mathbf x}_i \in \mathcal{S}_{\rm PDE}}{\bf G}_{{\rm new},i}^2 + \varepsilon_1\frac{1}{ |\mathcal{S}_{\rm IBs}|} \sum_{{\mathbf x}_i \in \mathcal{S}_{\rm IBs}} (\hat{\bf U}_{e,i} - {\bf U}_{e0,i})^2,
\end{equation}

The architecture of PINN-WE for solving conservative hyperbolic equations is shown in Fig. \ref{fig:PINNWE}.

We define the gradient-dependent weight as
\begin{equation}\label{eq:6}
  \lambda = \frac{1}{ \varepsilon_2 (|\nabla \cdot \vec{u}| - \nabla\cdot \vec{u})+1}.
\end{equation}

where $\vec{u}$ is the velocity field. 
In \cite{vonneumann1950method}, artificial viscosity is added into the  governing equations according to the velocity divergence  $\nabla \cdot \vec{u}$  considering that the field is compressed when $\nabla \cdot \vec{u} < 0$.
Here, we only use the velocity divergence to detect shocks and apply  $\lambda$ into the equations without adding any numerical dissipation to the equations. As $\lambda$ is constant positive, so it will not influence the exact solution  of the equations.

Fig.\ref{fig:PINNWE} shows the architecture of the proposed method to solve the given PDE (\ref{eq:pde}). We evaluate our proposal in the following numerical examples.

\section{Numerical examples }
\footnote{Code can be found in $https://github.com/bfly123/PINN\_WE $}

\subsection{Burgers equation}
We first resolve Burgers equation (\ref{eq:burgers}) using the new PINN-WE method with a totally same setting with section 2.2.

\begin{figure}
  \centering
  \subfigure[Loss history]{
\includegraphics[width = 9cm]{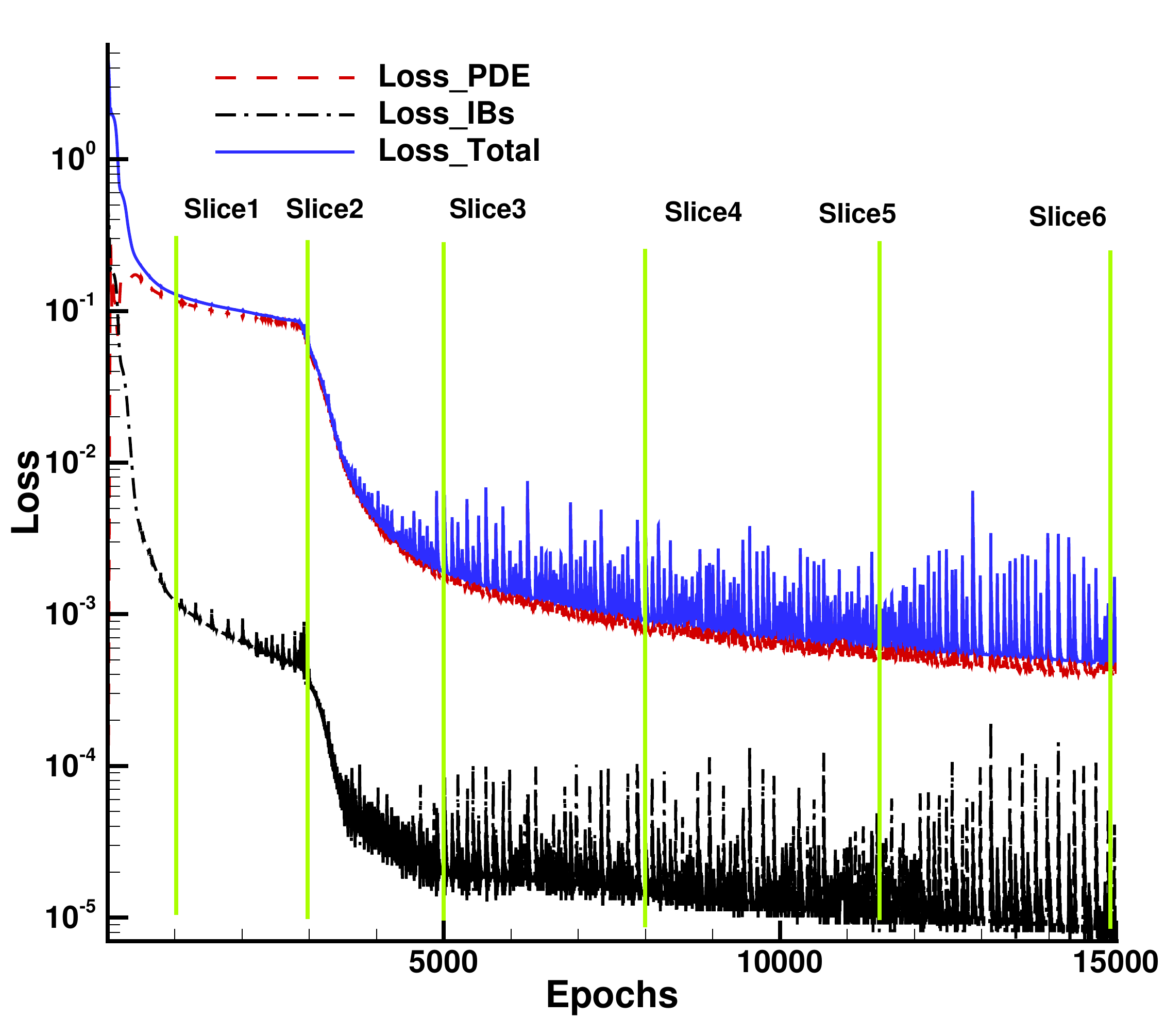}}
  \subfigure[$u$ at $t=1$ at different epochs]{
\includegraphics[width = 9cm]{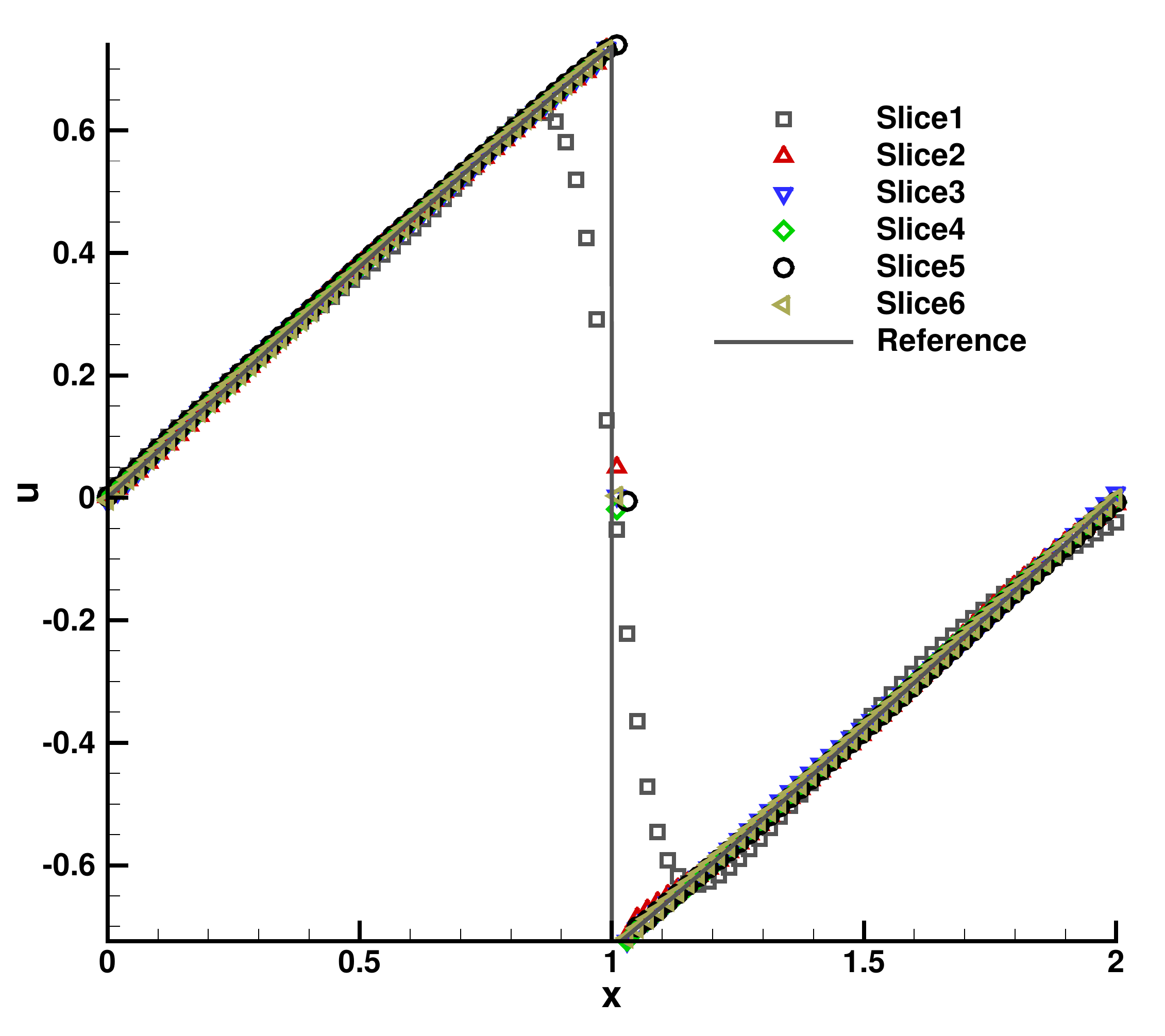}}
  \subfigure[Residual of $x$ at $t=1$ at different epochs]{
\includegraphics[width = 9cm]{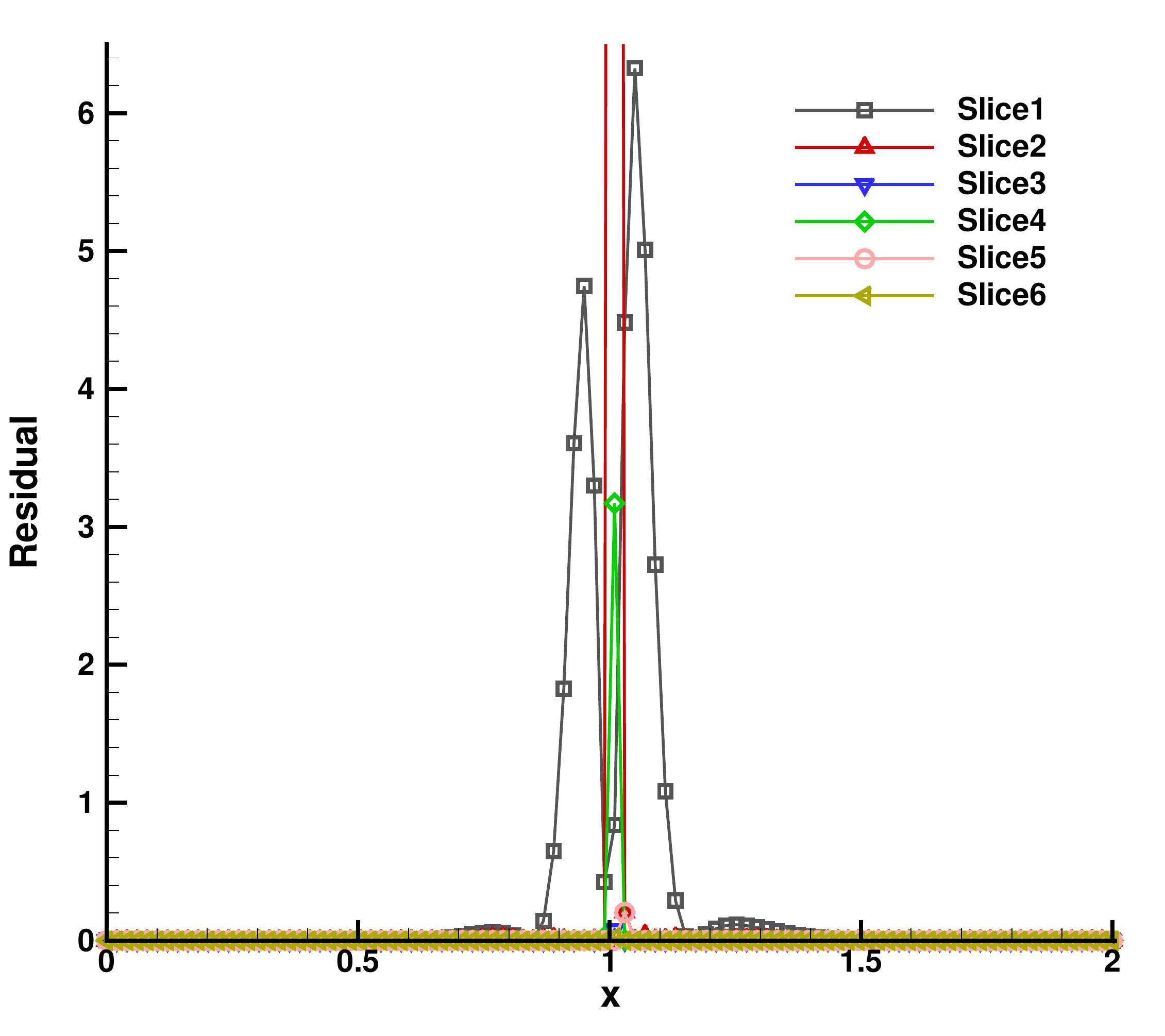}}
  \subfigure[Weight $\lambda$ of $x$ at $t=1$ at different epochs]{
\includegraphics[width = 9cm]{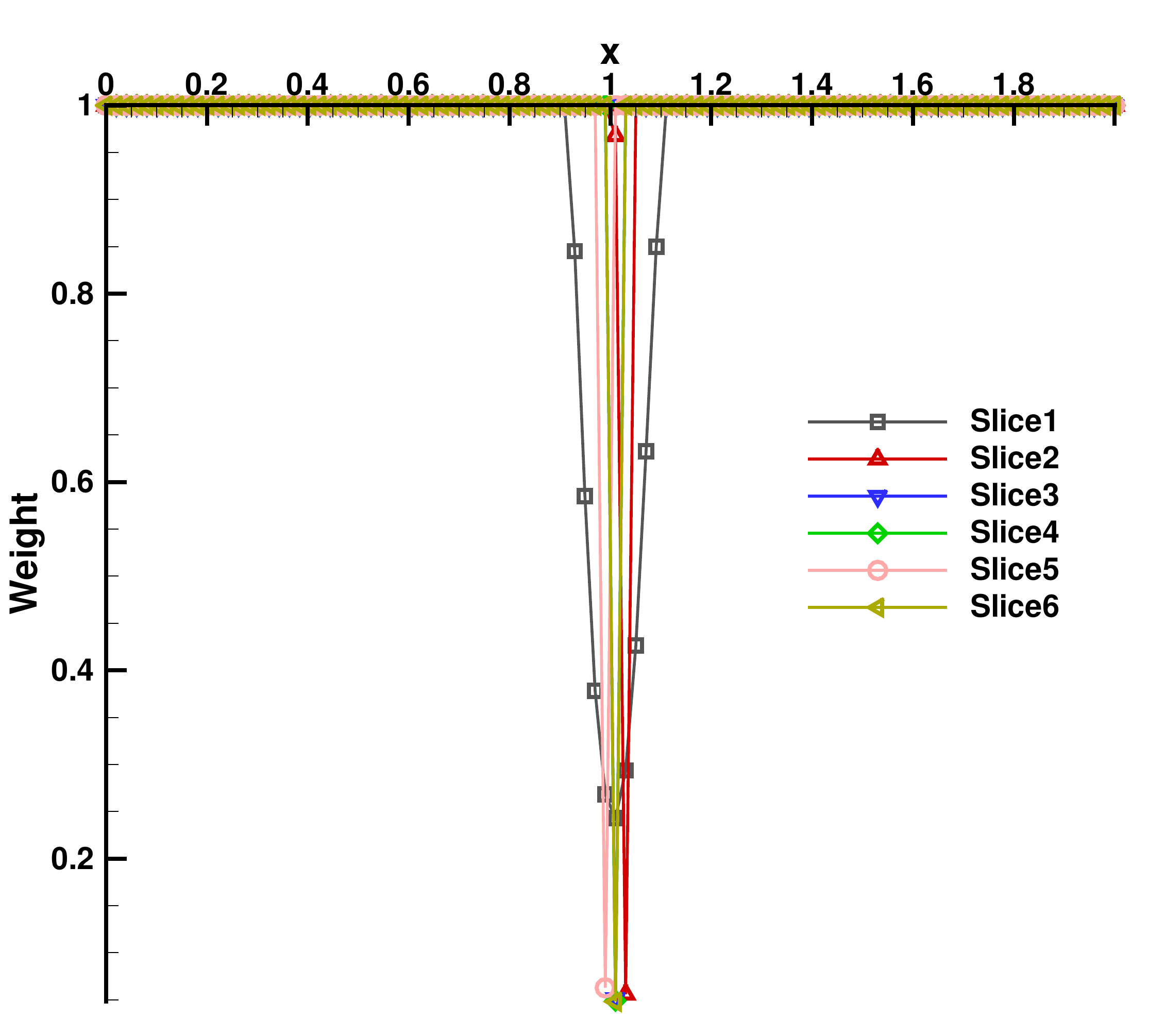}}
\caption{Results of Burgers equation with PINN-WE. Same computational setting with Fig.\ref{fig:burgers_1} }\label{fig:burgers_2}
\end{figure}

In Burgers equation the velocity is $u$ itself. So the weight in ${\bf G}_{\rm new}$ is 
\begin{equation}
  \lambda = \frac{1}{ \varepsilon_2 (|\frac{\partial u}{\partial x}| -  \frac{\partial u}{\partial x})+1}.
\end{equation}
Fig.\ref{fig:burgers_2} gives the loss history of epochs and  $u$, the residual and $\lambda$ as functions of $x$ at $t=1$ at different training epochs (1000,3000,5000,8000,11500 and 14931 (which has the minimum total loss)). Compared to the result in Fig.\ref{fig:burgers_1}, in the first 1000 epochs (Slice 1), PINN-WE has a similar result with traditional PINN as the gradient of transition-point is not large enough to cause problems. However, after 3000 epochs, PINN-WE can effectively deduce the total loss. All the transition-points are compressed into smooth region besides the one $u=0$ which has zero velocity. As shown in Fig.\ref{fig:burgers_2}.d, after 3000 epochs, all the weights change to nearly $1$ besides the central one which has large gradient.    

%


\subsection{Euler equations}
Then we consider the following 1D and 2D Euler equations in conservative forms: 
\begin{equation}
\frac{\partial {\bf U}}{\partial t} + \nabla \cdot {\bf F}  = 0.
\end{equation}
For the 1D case,
\begin{equation}
  {\bf{U}} = \left(\begin{aligned}
    &\rho\\
    &\rho u\\ 
    &E
  \end{aligned}\right),
  {\bf F} = \left(\begin{aligned}
    & \rho u\\
    & \rho u^2 + p\\ 
  &  u(E + p)
  \end{aligned}\right).
\end{equation}
For 2D cases, ${\bf F}=({\bf F}_1,{\bf F}_2)$, where
\begin{equation}
  \bf{U} = \left(\begin{aligned}
    &\rho\\
    &\rho u\\ 
    &\rho v\\ 
    &E
  \end{aligned}\right),
  F_1= \left(\begin{aligned}
    & \rho u\\
    & \rho u^2 + p\\ 
    & \rho uv\\ 
  &  u(E + p)
  \end{aligned}\right),
F_2= \left(\begin{aligned}
    & \rho v\\
    & \rho uv\\ 
    & \rho v^2 +p\\ 
  &  v(E + p)
  \end{aligned}\right).
\end{equation}
Above, $\rho$ is the density, $u$ is the velocity, $(u,v)$ is the 2D velocity vector, $p$ is the pressure, and $E$ is the total energy. To close the equations, we use the following equation of state for ideal gas:
\begin{equation}
   E = \frac{1}{2}\rho u^2 + \frac{p}{\gamma -1},
\end{equation}
where $\gamma = 1.4$ is the specific heat ratio.

We consider forward problems with PINN to solve weighted Euler equations ${\bf G}_{\rm new} =0$ directly without more data besides the IBs. In 1D cases, we consider the classical Sod and Lax problems. Then, we consider a 2D Riemann problem and a more complex moving shock problem. We compare all the results with those obtained from traditional high-order finite differential methods, namely, fifth-order WENO-Z method in spatial discretization \cite{borges2008improved} and third-order Runge-Kutta method \cite{shu1988efficient} for time integration.
\subsubsection{Sod problem}
The Sod problem has been extensively studied. It is a 1D Riemann problem with the initial constant states in a tube with unit length formulated as
\begin{equation}
  (\rho,u,p) = \left\{  \begin{aligned} 
    &(1,0,1), \quad  &{\rm if} \quad 0\le x \le 0.5,\\
    &         (0.1,0,0.125), & \quad {\rm if} \quad 0.5< x \le 1.\\
  \end{aligned}\right.
\end{equation}

\begin{figure}
  \centering
\includegraphics[width = 9cm]{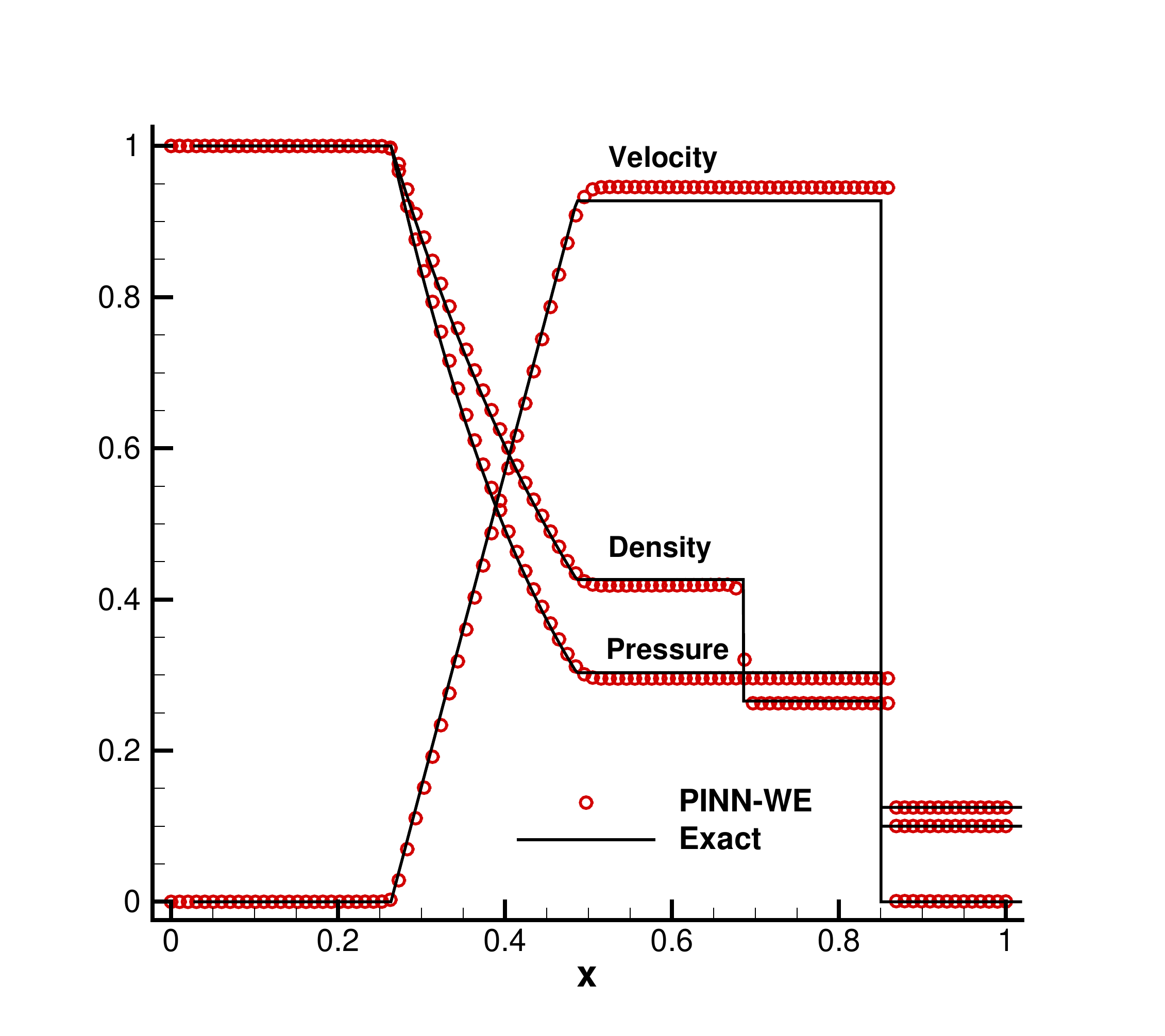}
\includegraphics[width = 9cm]{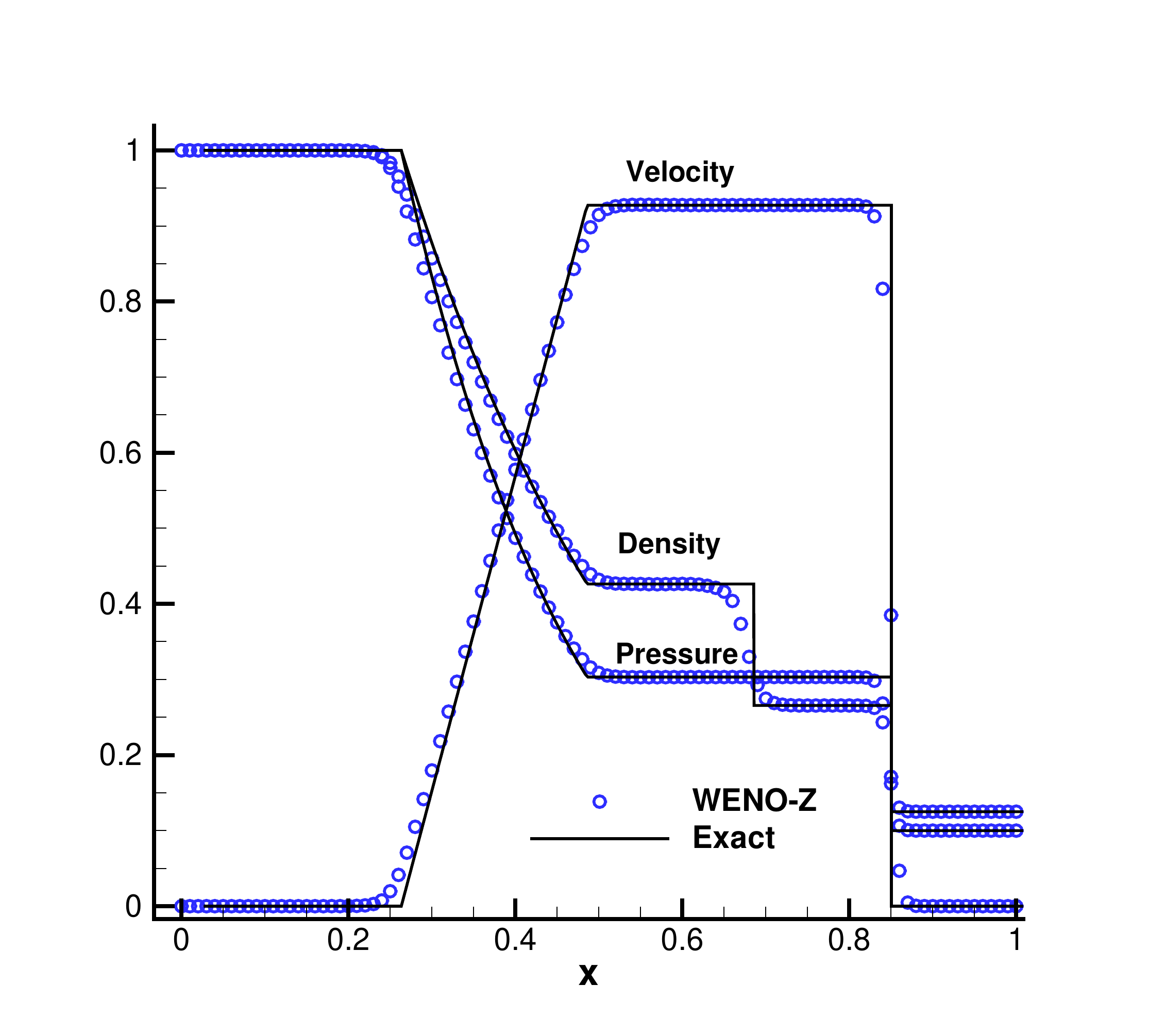}
\caption{Evaluation of Sod problem at $t=0.2$.}\label{fig:sod1}
\end{figure}

\begin{figure}
  \centering
\includegraphics[width = 15cm]{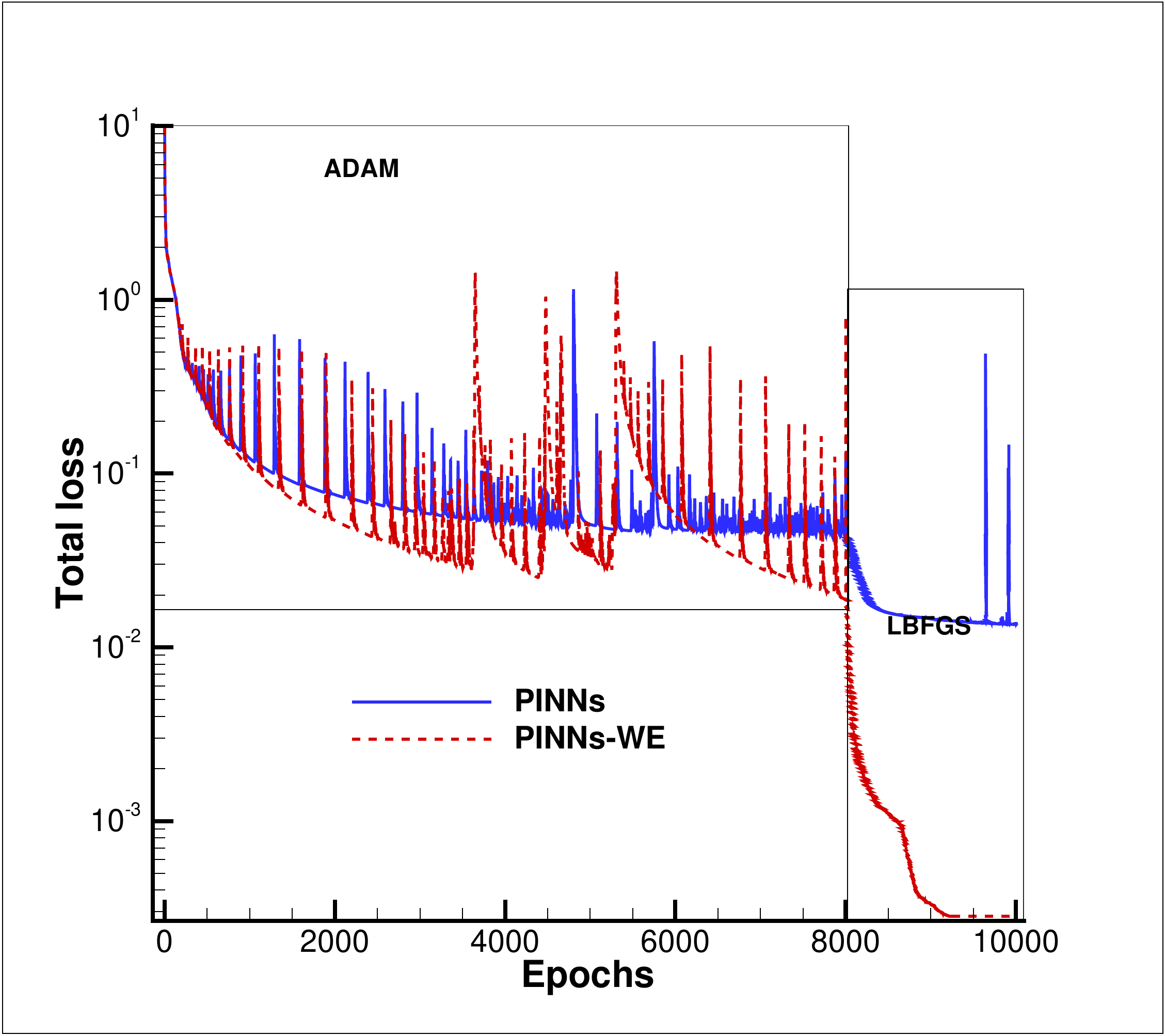}
\caption{Training loss evolution for Sod problem.} 
\end{figure}

\begin{figure}
  \centering
\includegraphics[width = 6cm]{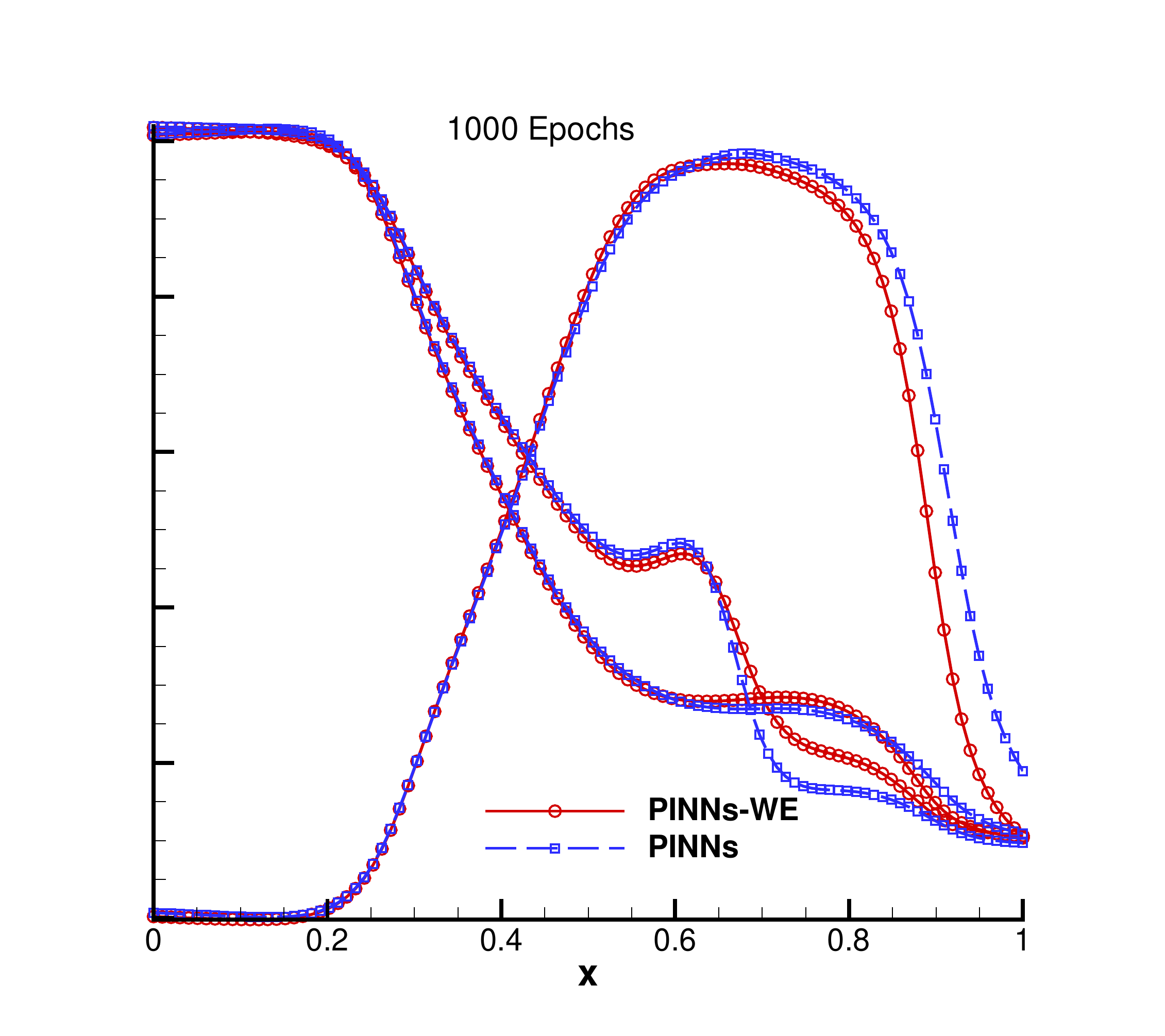}
\includegraphics[width = 6cm]{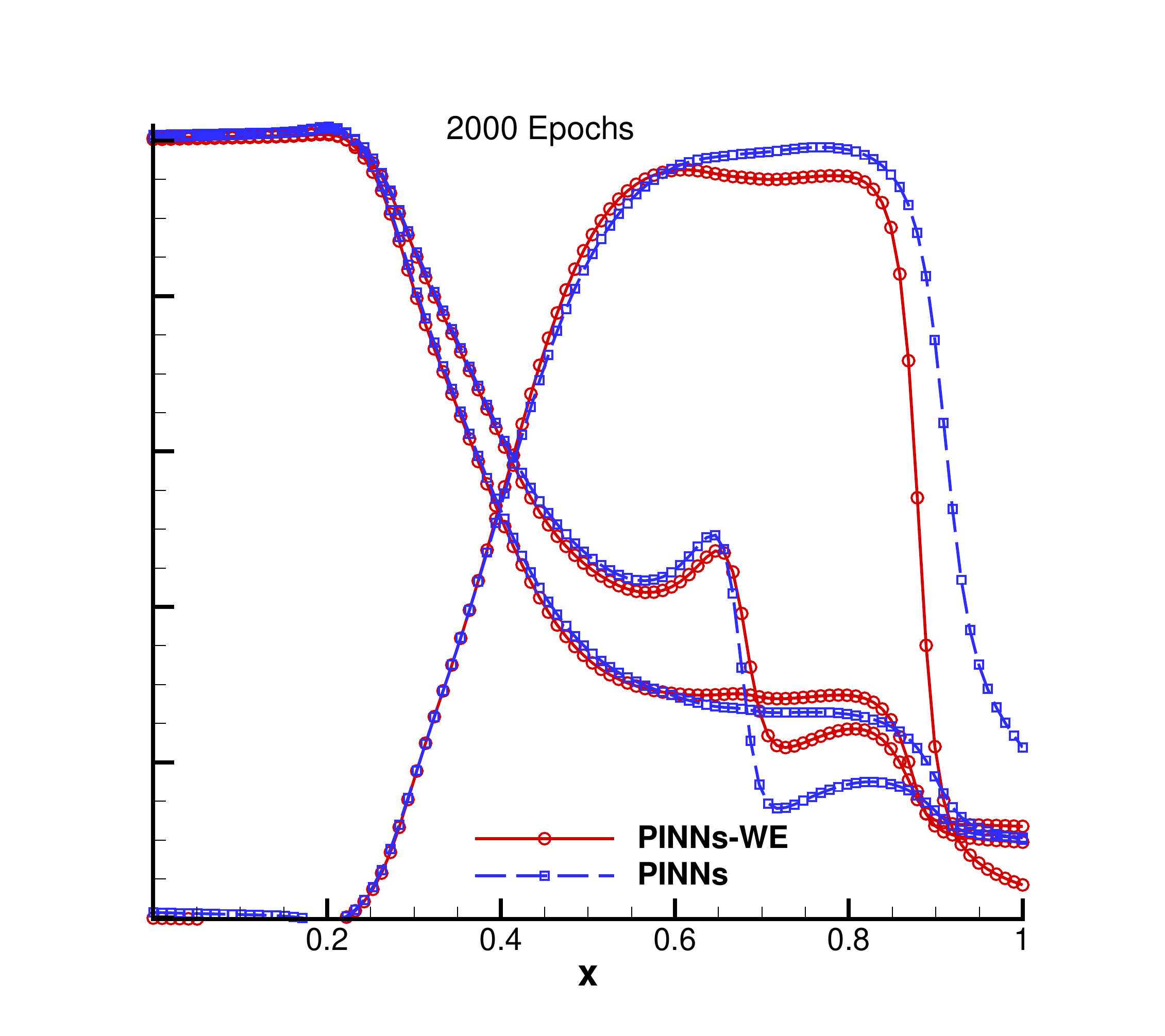}
\includegraphics[width = 6cm]{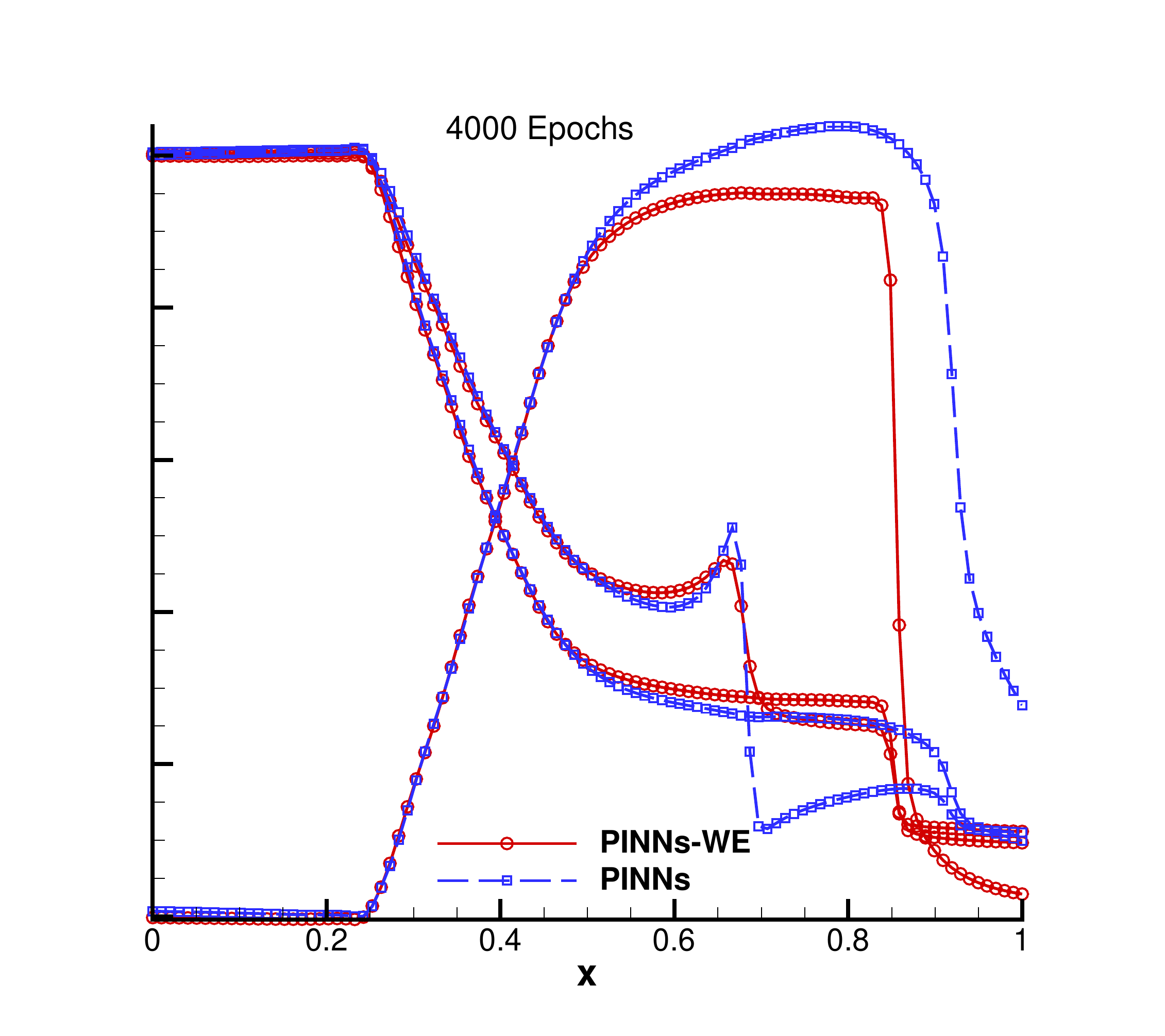}
\includegraphics[width = 6cm]{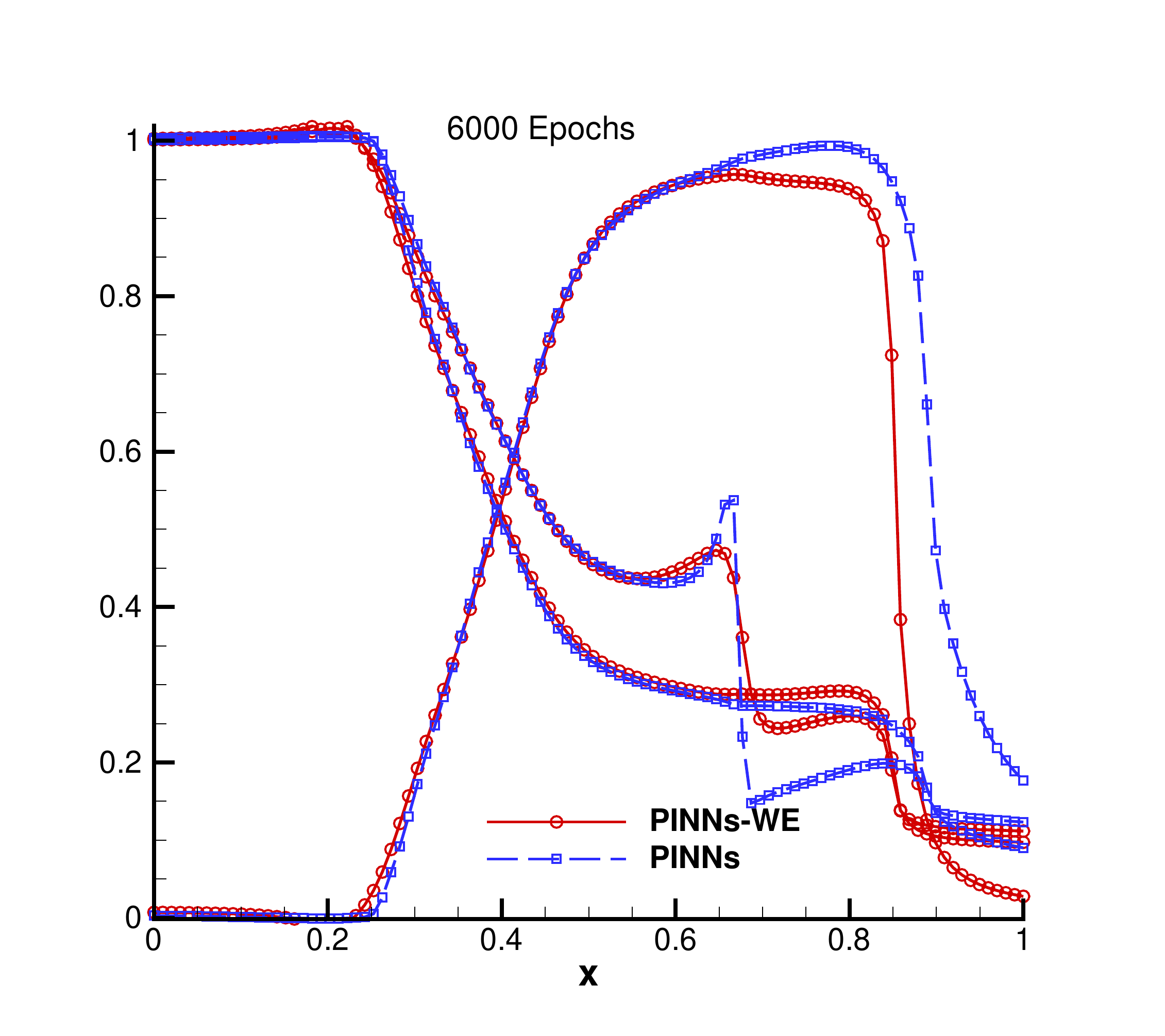}
\includegraphics[width = 6cm]{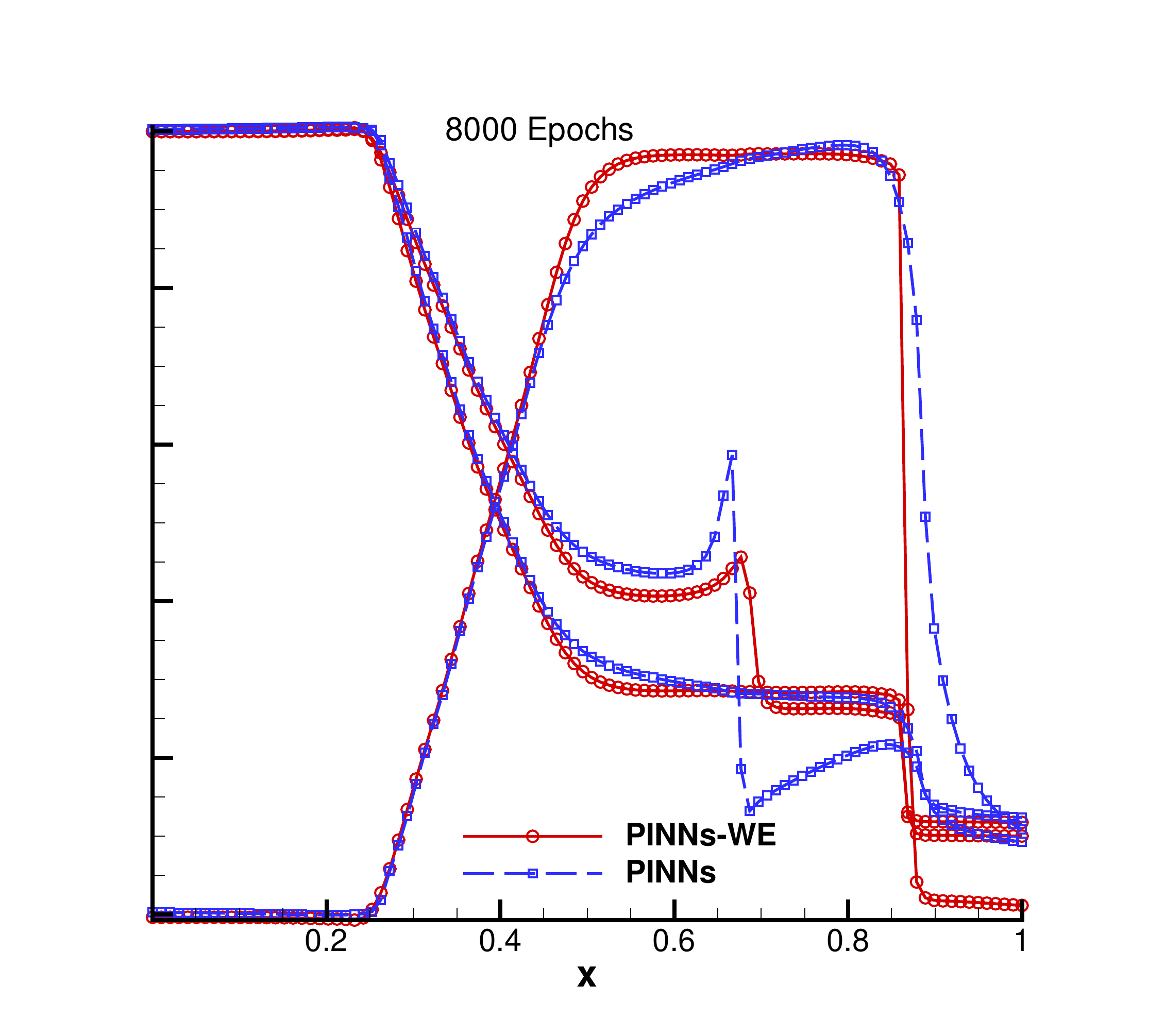}
\includegraphics[width = 6cm]{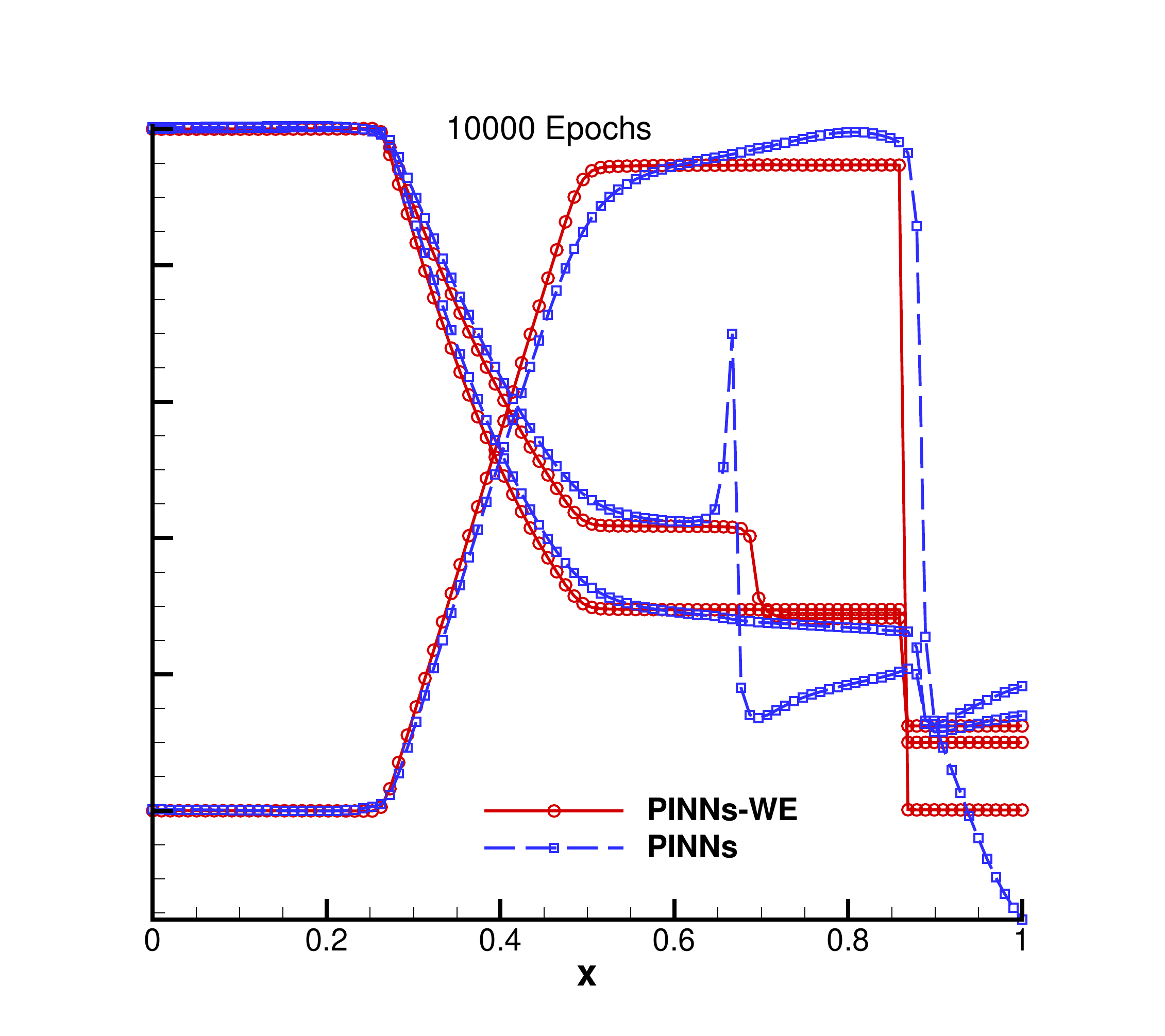}
\caption{Results at different training epochs for Sod problem.} 
\end{figure}

We used an NN with 7 hidden layers and 50 neurons per layer. Then, we randomly selected 10,000 residual points from a uniform mesh of $100\times 200$ in the $X \times T$ space. The number of IBs points was $1000$.  After training,  we constructed a test set with $100$ uniform points in $x\in [0,1]$ at final time $t =0.2$. 

We first compared the results with the traditional high-order WENO-Z method for $100$ cells in space. Fig. \ref{fig:sod1} shows that  
PINN-WE achieves similar and even better results compared with the WENO-Z method. Especially in capturing shock waves, no transition-points occur inside the shock because no dissipation is introduced into the equations for PINN-WE. 

Fig. 3 shows the training loss of PINN and PINN-WE, and Fig. 4 shows results at different training epochs to evaluate the training evolution. In the first 1000 epochs, there is a small difference between PINN and PINN-WE because no strong compression occurs. Then, PINN straggles when training a shock, while PINN-WE can easily avoid this region. With the loss decreasing at smooth regions, a sharp shock appears with the transition-points compressed by the left and right smooth regions. 

\subsubsection{Lax problem}
The Lax problem is also a Riemann problem that contains a strong shock and strong contact. Its initial condition is given by 
\begin{equation}
  (\rho,u,p) = \left\{  \begin{aligned} 
    &(0.445,0.698,3.528), \quad  &{\text if} \quad 0\le x \le 0.5,\\
    &         (0.5,0,0.571), & \quad {\text if} \quad 0.5< x \le 1.\\
  \end{aligned}\right.
\end{equation}
We used the same NN as for the Sod problem and randomly selected 50,000 interior points from a uniform $1000 \times 5000$ mesh in the $X \times T$ space. We considered 1000 initial points. This problem was more difficult to train than the Sod problem, and the total loss was 0.01 after training for 20,000 epochs. We compared the results with those obtained from the high-order WENO-Z method. Fig. 5 shows that PINN-WE correctly captures shocks.    
\begin{figure}
  \centering
\includegraphics[width = 9cm]{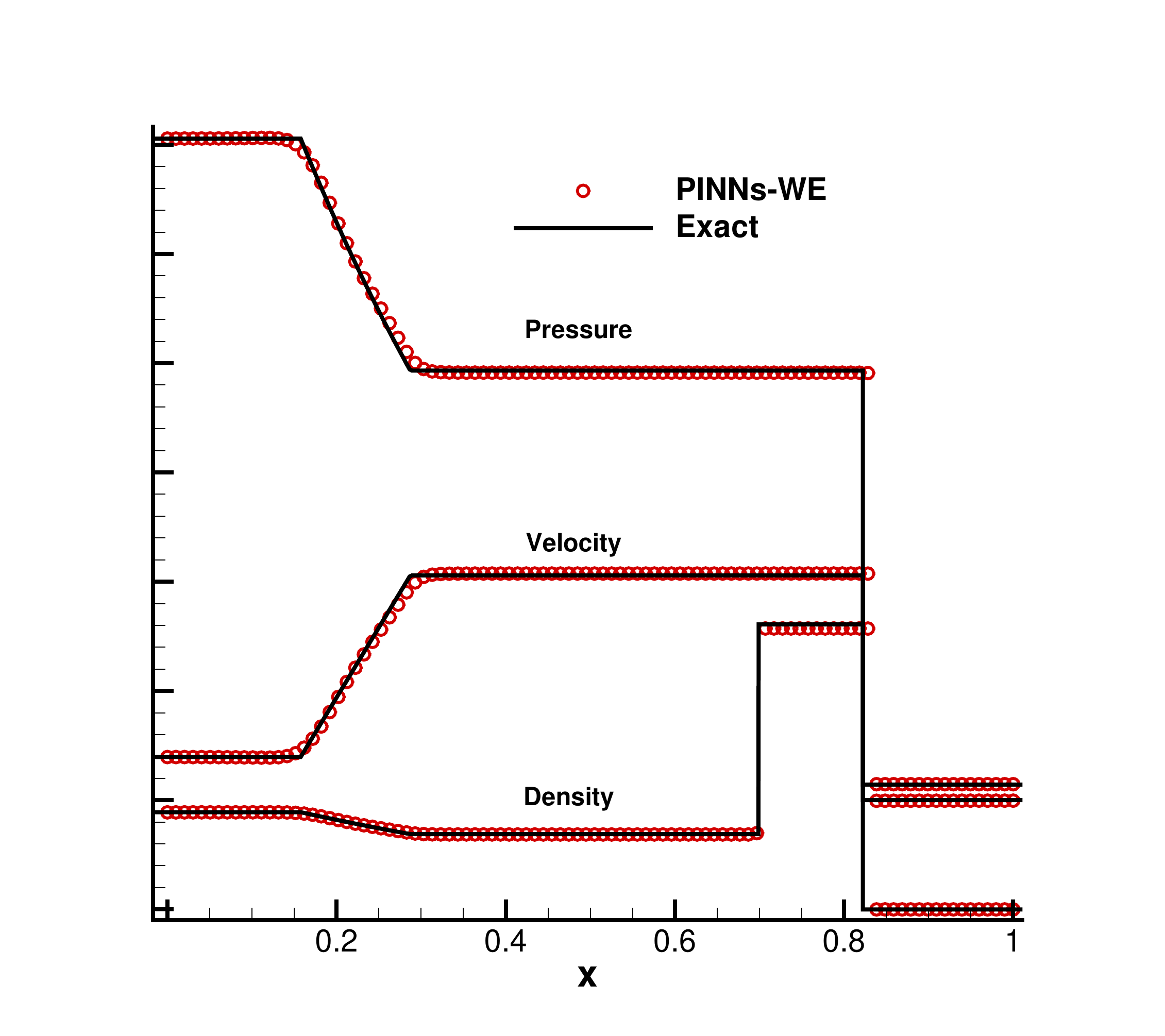}
\includegraphics[width = 9cm]{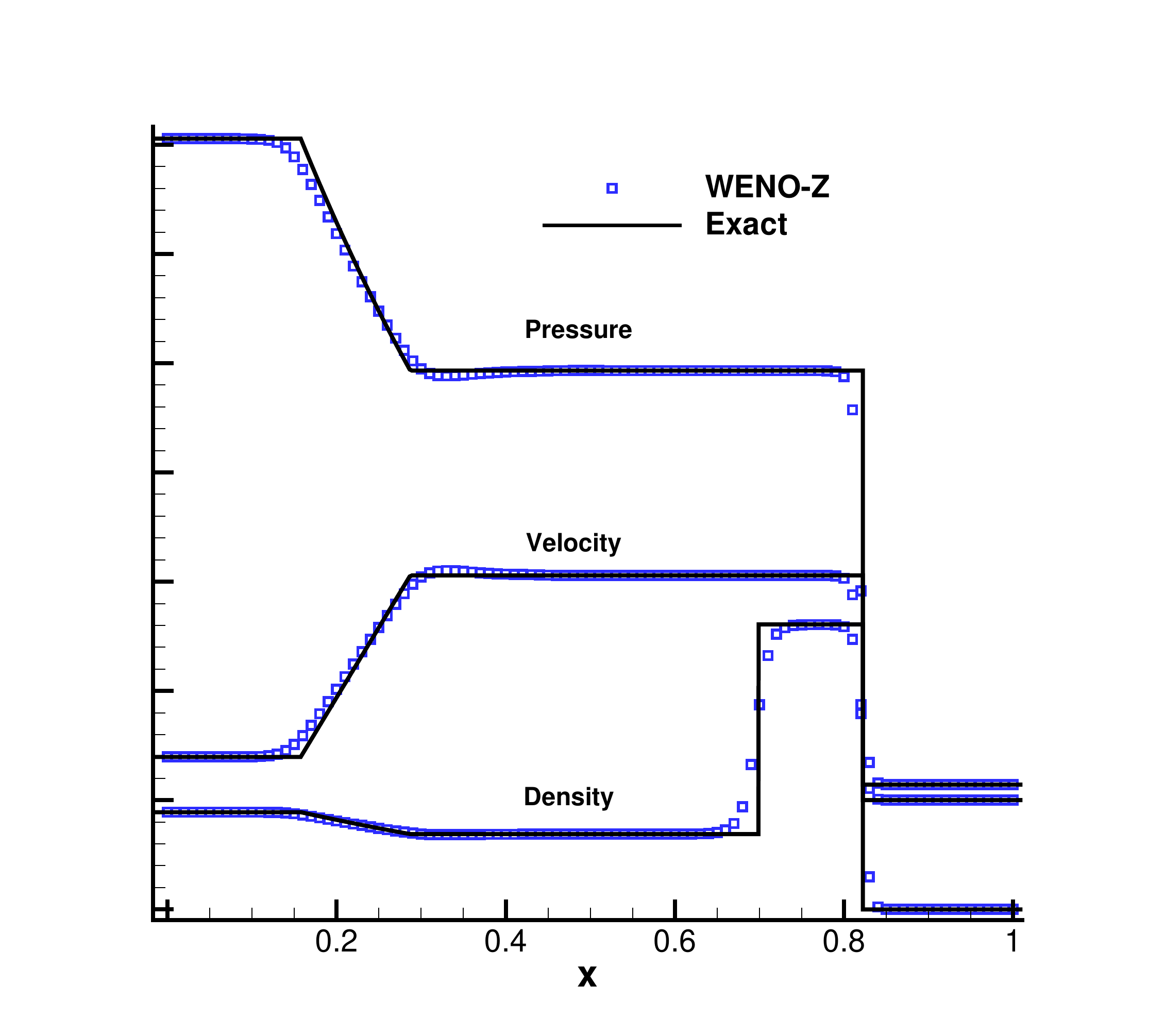}
\caption{Results and loss for Lax problem.} 
\end{figure}

\subsubsection{2D Riemann problem}

Then, we considered a 2D problem with strong contact discontinuities. The basic settings were retrieved from case 8 in \cite{kurganov2002solution}. The initial condition is given by
\begin{equation}
  (\rho,u,v,p) = \left\{
   \begin{aligned}
    & (1,-0.75,0.5,1)  &\quad \text{if}  &\quad  0\le x \le 0.5, 0\le y \le 0.5,\\
    & (2,0.75,0.5,1)   &\quad  \text{if} &\quad  0\le x \le 0.5, 0.5< y \le 1,\\
    & (3,-0.75,-0.5,1)  &\quad \text{if} &\quad  0.5 < x \le 1, 0\le y \le 0.5,\\
    & (1,0.75,-0.5,1)  &\quad \text{if} &\quad  0.5 \le x \le 0.5, 0.5 < y \le 1.\\
   \end{aligned}
   \right.
  \end{equation}

We used an NN with 6 hidden layers and 60 neurons per layer. The training points were obtained by Latin hypercube sampling, with 200,000 interior points in the $T\times X \times Y$ space and 10,000 initial points in the $X \times Y$ space.
The final training loss was $0.009$.  We evaluated the model with meshes of $100\times 100$ and $400 \times 400$ in the $X \times Y$ space at instants $0.2$ and $0.4$, respectively. Then, we compared the results with those obtained from the WENO-Z scheme computed in a $100\times 100$ mesh in the $X\times Y$ space. The corresponding results are shown in Fig. 6.   

\begin{figure}
  \centering
  \subfigure[Density PINN-WE]{
\includegraphics[width = 9cm]{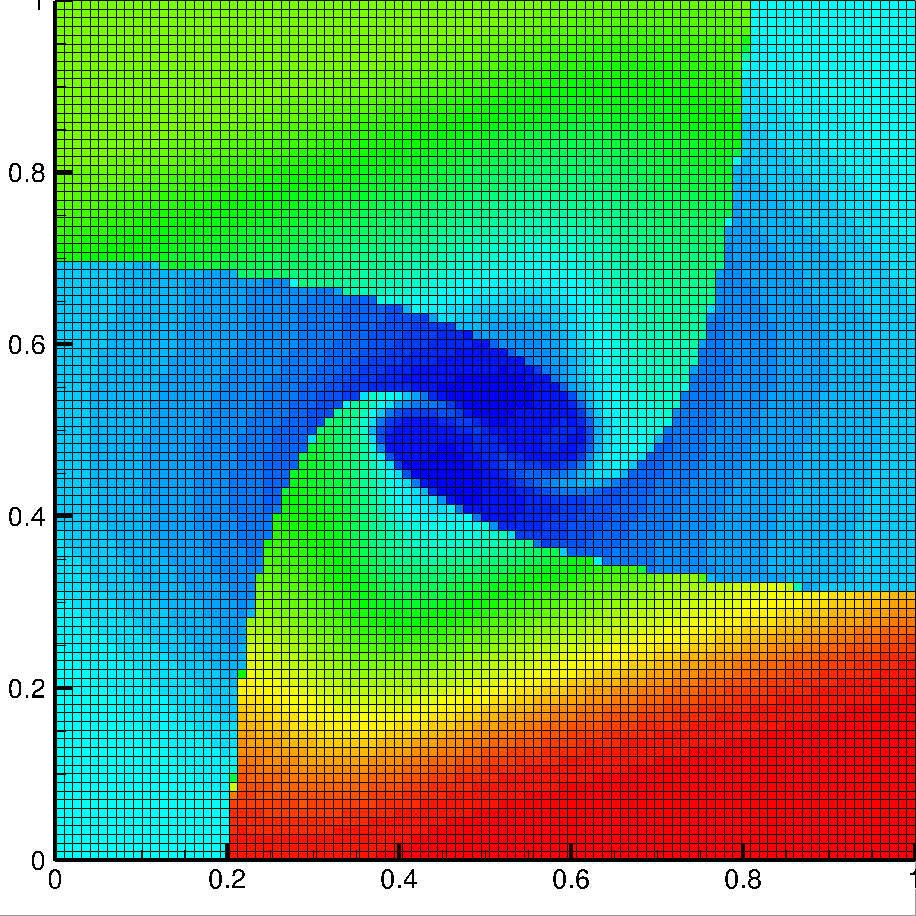}}
  \subfigure[Density WENO-Z]{
\includegraphics[width = 9cm]{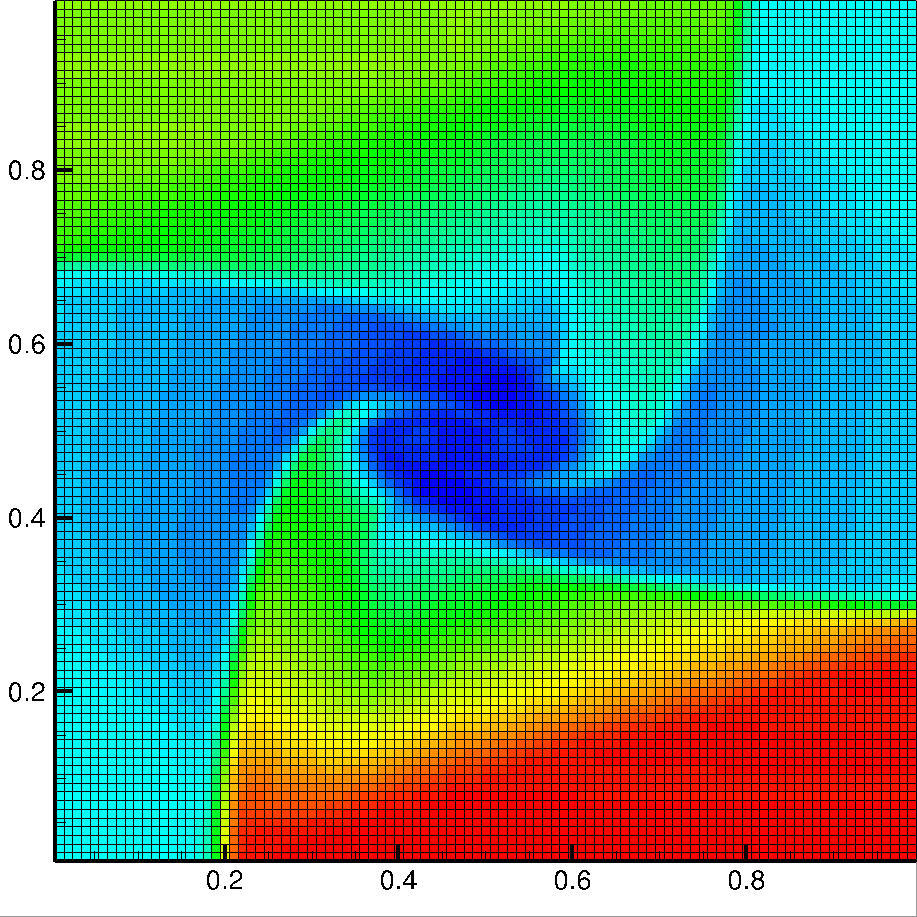}}
  \subfigure[U PINN-WE]{
\includegraphics[width = 9cm]{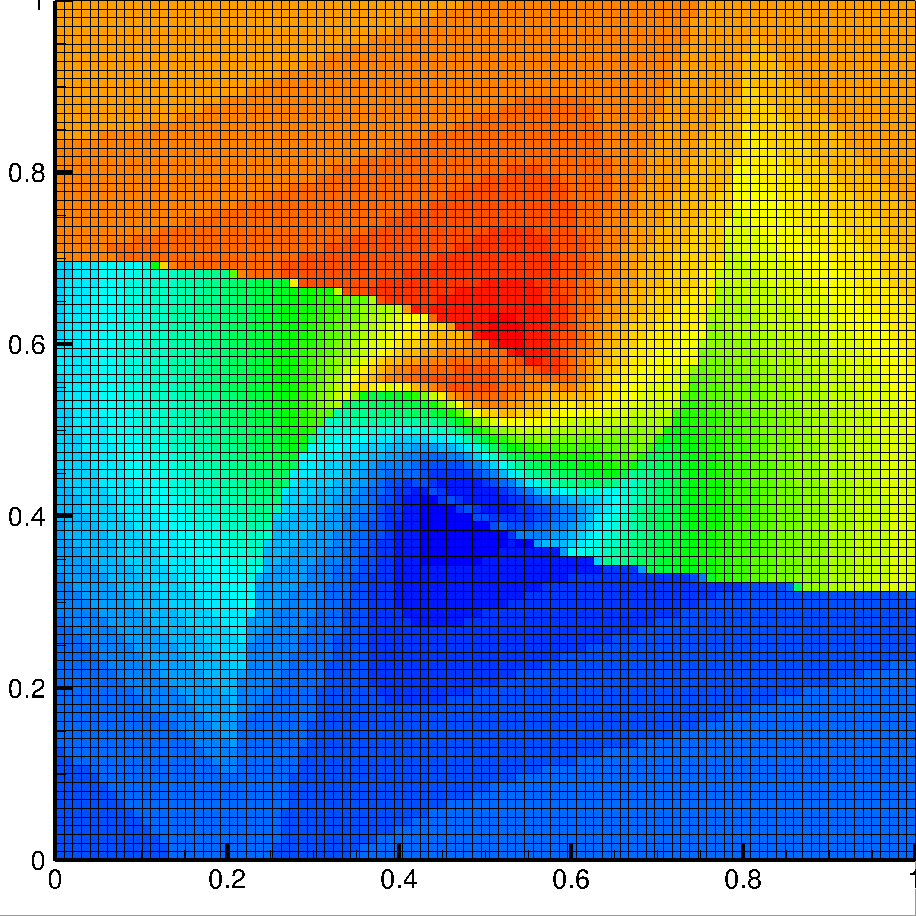}}
 \subfigure[U WENO-Z]{
\includegraphics[width = 9cm]{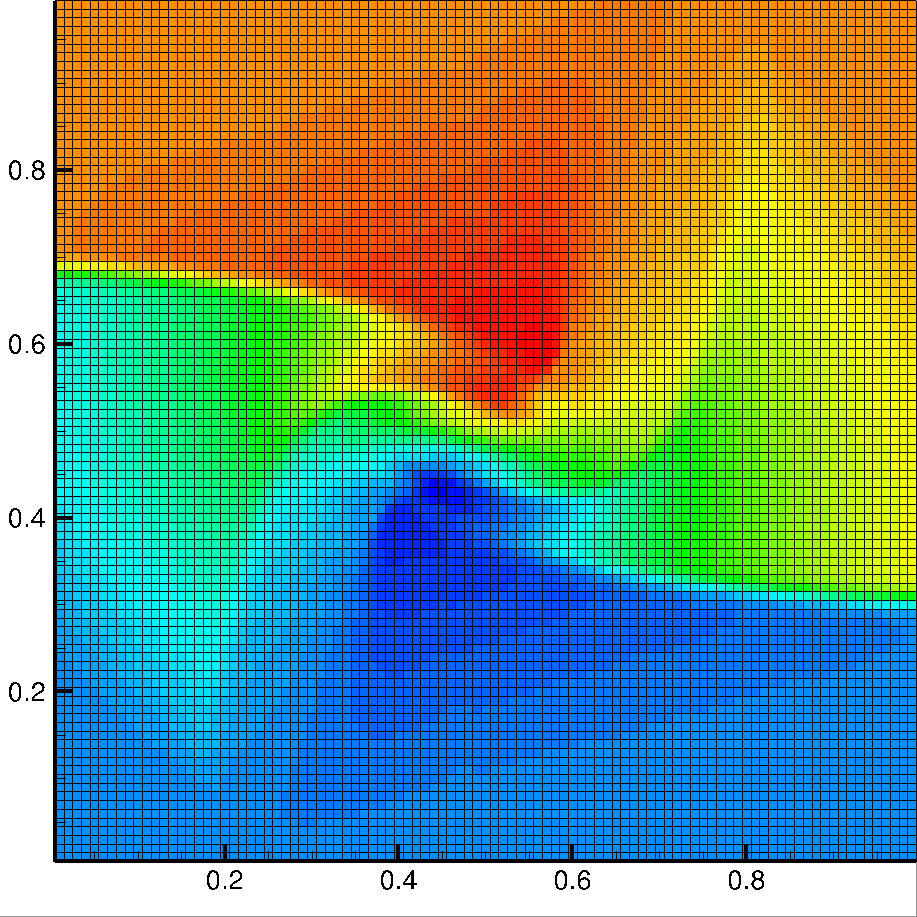}}
\caption{Results of 2D Riemann problem (part 1) with $100\times 100$ test points for PINN-WE and the same number of mesh grids for WENO-Z.}\label{fig:2driemann_1}
\end{figure}
\begin{figure}
  \subfigure[V PINN-WE]{
\includegraphics[width = 9cm]{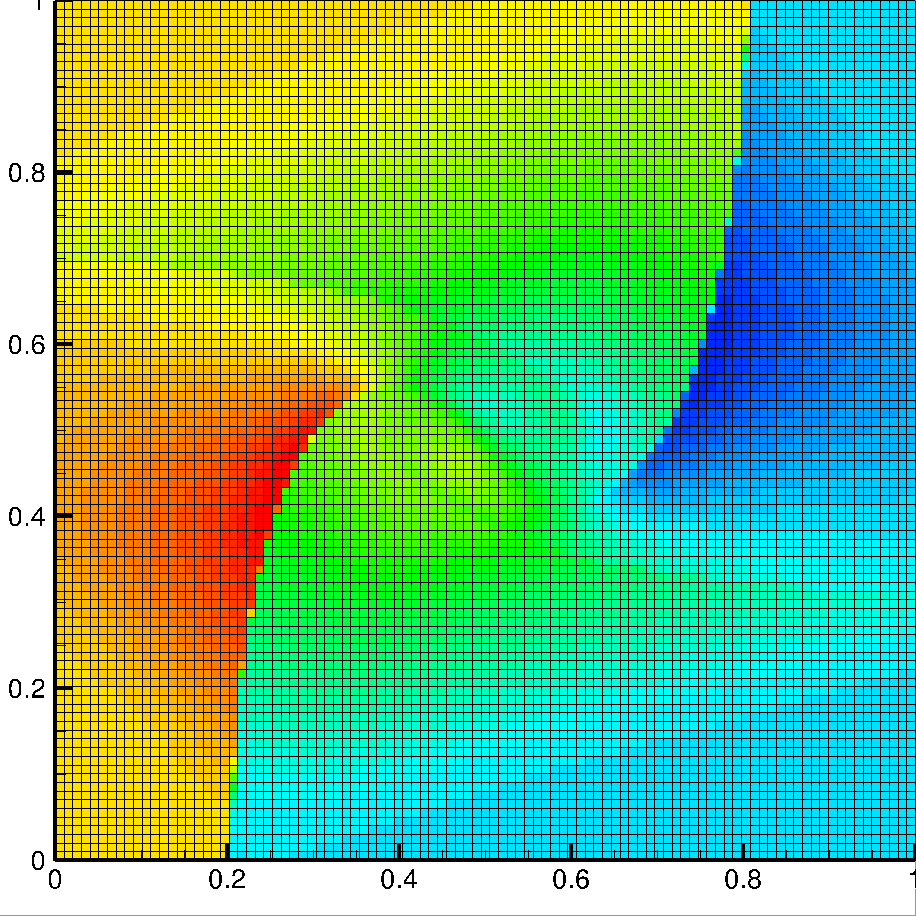}}
  \subfigure[V WENO-Z]{
\includegraphics[width = 9cm]{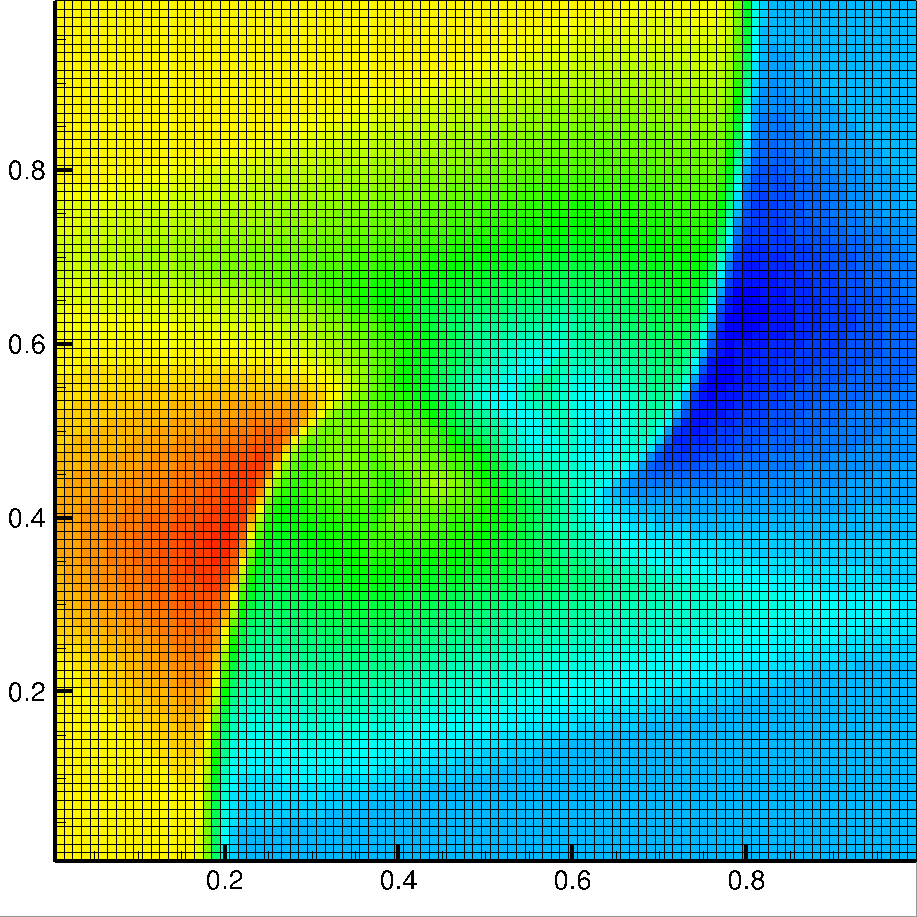}}
  \subfigure[Pressure PINN-WE]{
\includegraphics[width = 9cm]{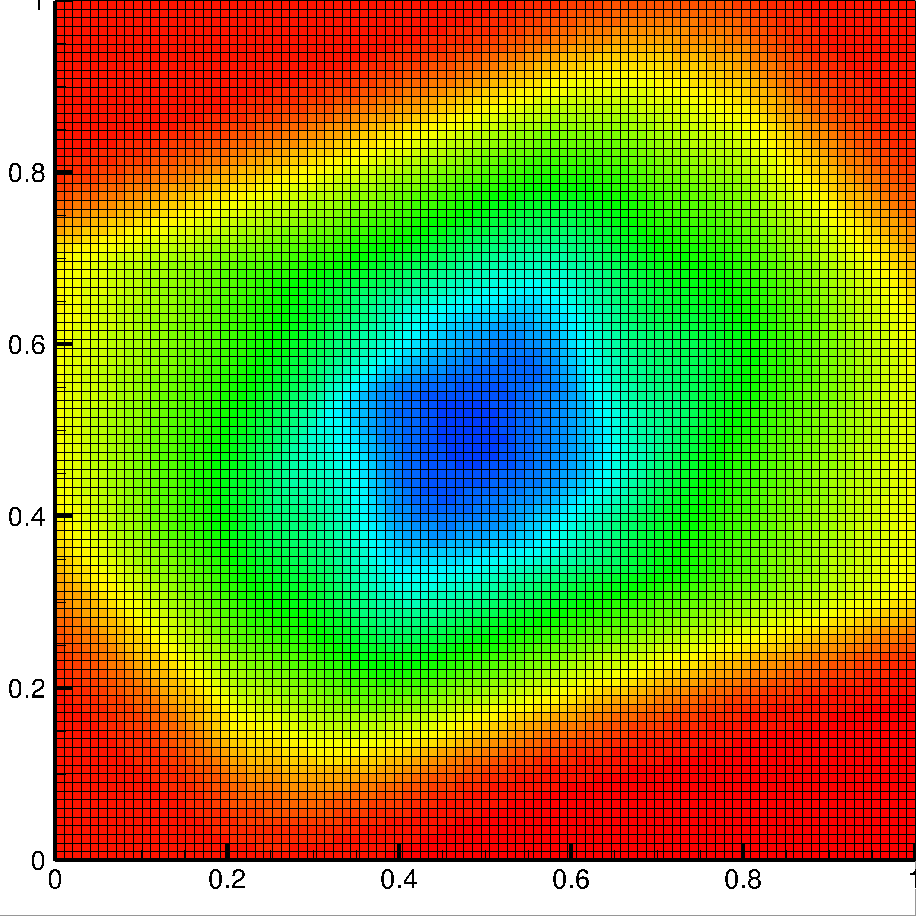}}
  \subfigure[Pressure WENO-Z ]{
\includegraphics[width = 9cm]{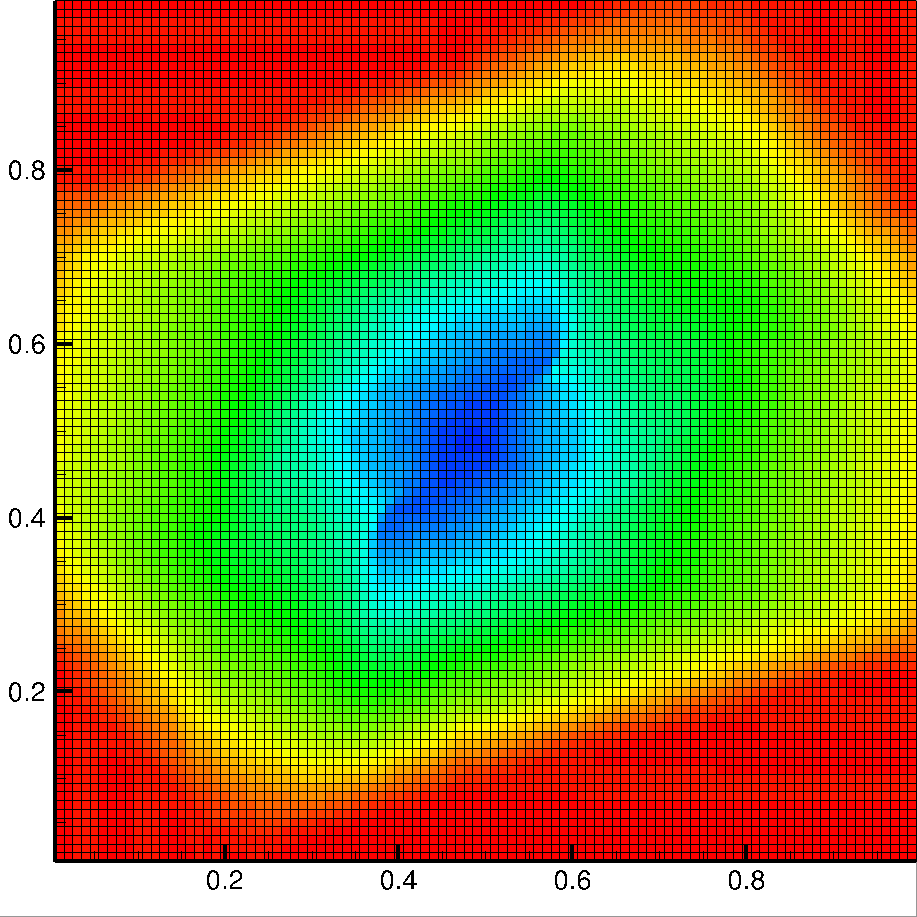}}
\caption{ Results of 2D Riemann problem (part 2) with $100\times 100$ test points for PINN-WE and the same number of mesh grids for WENO-Z.}\label{fig:2driemann_2}
\end{figure}

\begin{figure}
  \centering
  \subfigure[Density PINN-WE]{
\includegraphics[width = 9cm]{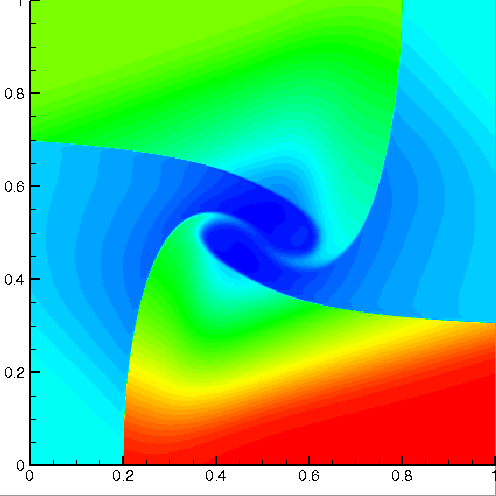}}
  \subfigure[Density WENO-Z]{
\includegraphics[width = 9cm]{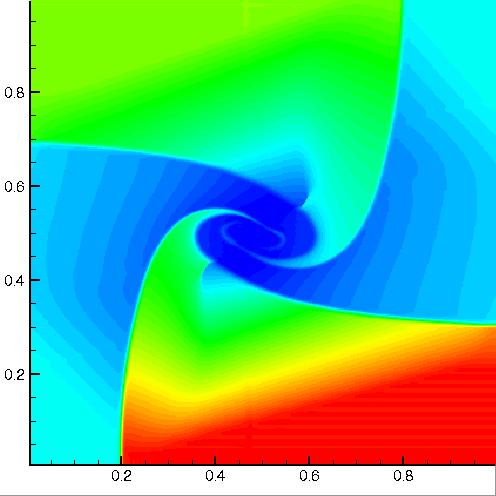}}
  \subfigure[U PINN-WE]{
\includegraphics[width = 9cm]{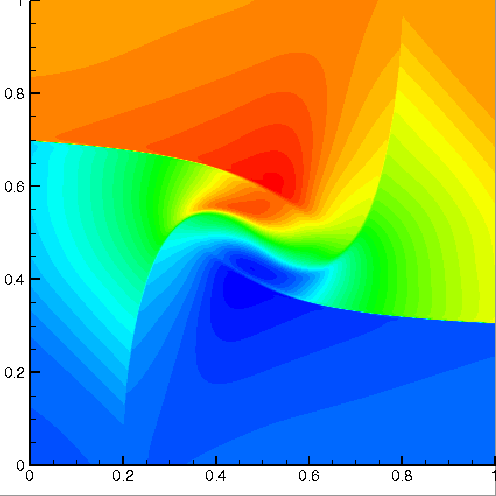}}
  \subfigure[U WENO-Z]{
\includegraphics[width = 9cm]{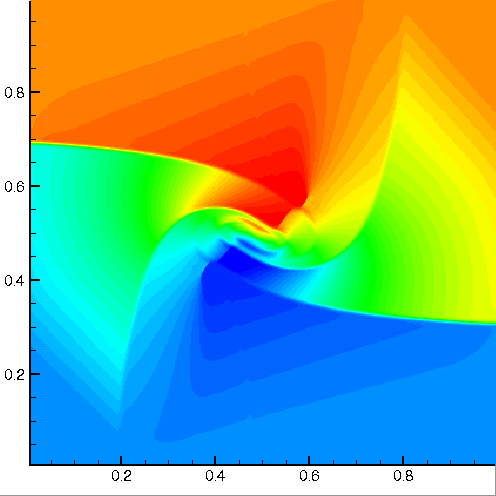}}
\caption{ Results of 2D Riemann problem (part 3) with $400\times 400$ test points for PINN-WE and the same number of mesh grids for WENO-Z.}\label{fig:2driemann_3}
\end{figure}
\begin{figure}
  \subfigure[V PINN-WE]{
\includegraphics[width = 9cm]{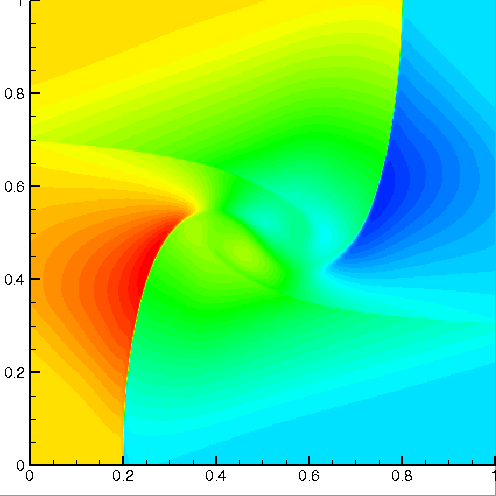}}
  \subfigure[V WENO-Z]{
\includegraphics[width = 9cm]{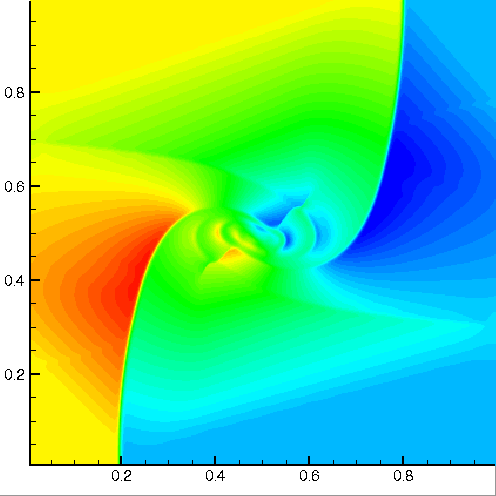}}
  \subfigure[Pressure PINN-WE]{
\includegraphics[width = 9cm]{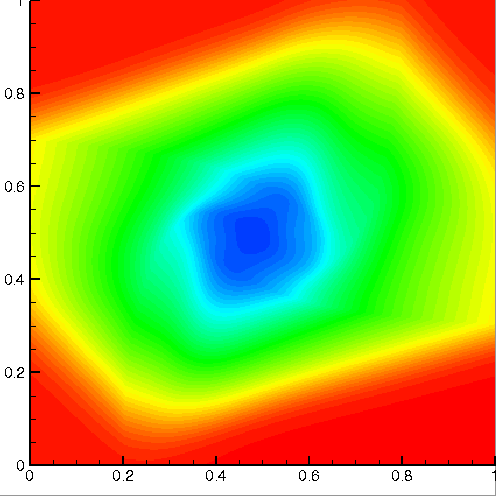}}
  \subfigure[Pressure WENO-Z]{
\includegraphics[width = 9cm]{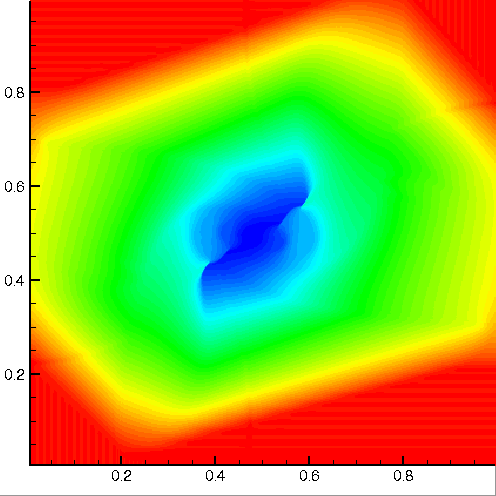}}
\caption{Results of 2D Riemann problem (part 4) with $400\times 400$ test points for PINN-WE and the same number of mesh grids for WENO-Z.}\label{fig:2driemann_4}
\end{figure}

We provide test results for $100 \times 100$ mesh points, which outnumber the training points (approximately 60) along each dimension. Comparison results are shown in Figs. \ref{fig:2driemann_1} and \ref{fig:2driemann_2} with the WENO-Z method in the same $100 \times 100$ mesh. For clarity, we show the original data without using the smoothing effect of the plotting software (Tecplot). The proposed method can capture contact discontinuities more sharply and nearly without transport and smoothing points, which are unavoidable by the traditional high-order method. Then, we increased the test points to a $400 \times 400$ mesh, which is substantially larger than training data along each dimension. Then, we performed an unfair comparison with the WENO-Z method in a $400 \times 400$ mesh. The computation of the detailed structure, especially in the middle region, is weaker in PINN-WE owing to the available training data, but discontinuities are still sharper than those computed by WENO-Z. These results illustrate the advantages of PINN-WE in high dimensions given its meshless feature.
 \subsubsection{Transonic flow through circular cylinder}

 Finally, we considered a 2D example with strong shock. A transonic flow with Mach number 0.728 passed through a stationary circular cylinder.
 A similar problem and its detailed analysis can be found in \cite{mo2018immersed}. The initial condition was given by a uniform flow with $(\rho,u,v,p) = (2,112,1.028,0,3.011)$. The computational domain was $T=[0,0.4]$,$X=[0,1.5]$ and $Y=[0,2]$. The center of the cylinder of radius  $0.25$ was located at $(1,1)$.
  We used an NN with 7 hidden layers and 90 neurons per layer. The collocations were obtained from Latin hypercube sampling in 3D domain $T\times X \times Y$ for $300,000$ points.
 In addition, $15,000$ boundary points were randomly sampled on the cylinder, and $15,000$ initial points were obtained from Latin hypercube sampling.
   The total loss was $0.028$ after 2000 steps of optimization with the L-BFGS algorithm.

   Figs. 10--14 show the comparison results with the WENO-Z method. Sharp and similar but smoother results are obtained from PINN-WE compared with WENO-Z. Moreover, our proposal can clearly capture the vertex behind the cylinder. However, limited by the number of the collocation samples owing to the hardware we used, we could not further decrease the loss to obtain a more exact result of small structures. Nevertheless, PINN-WE seems promising for computing complex transonic and supersonic flows.

 \begin{figure}
\includegraphics[width = 9cm]{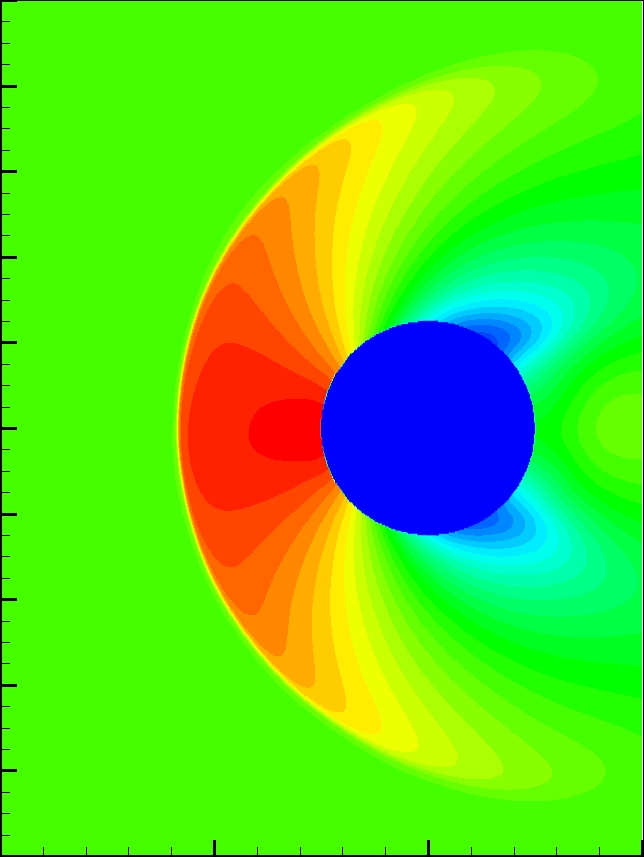}
\includegraphics[width = 9cm]{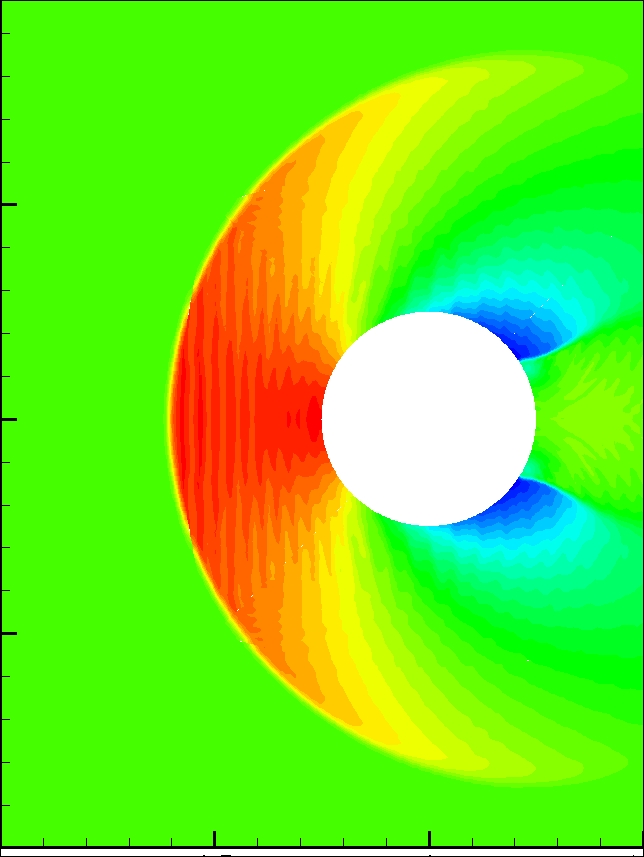}
\caption{Resulting pressure for transonic flow through circular cylinder using proposed PINN-WE (left) and WENO-Z method (right).} 
\end{figure}

 \begin{figure}
\includegraphics[width = 9cm]{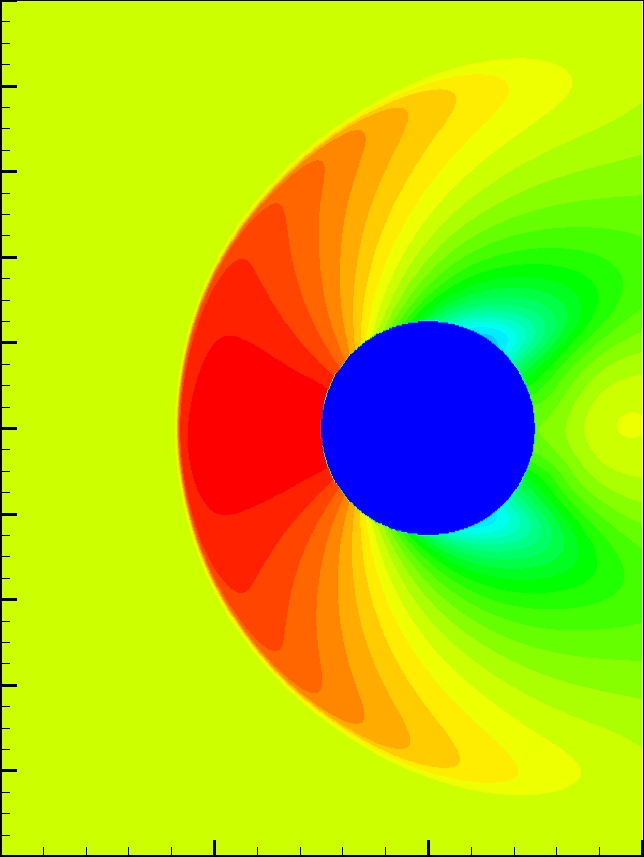}
\includegraphics[width = 9cm]{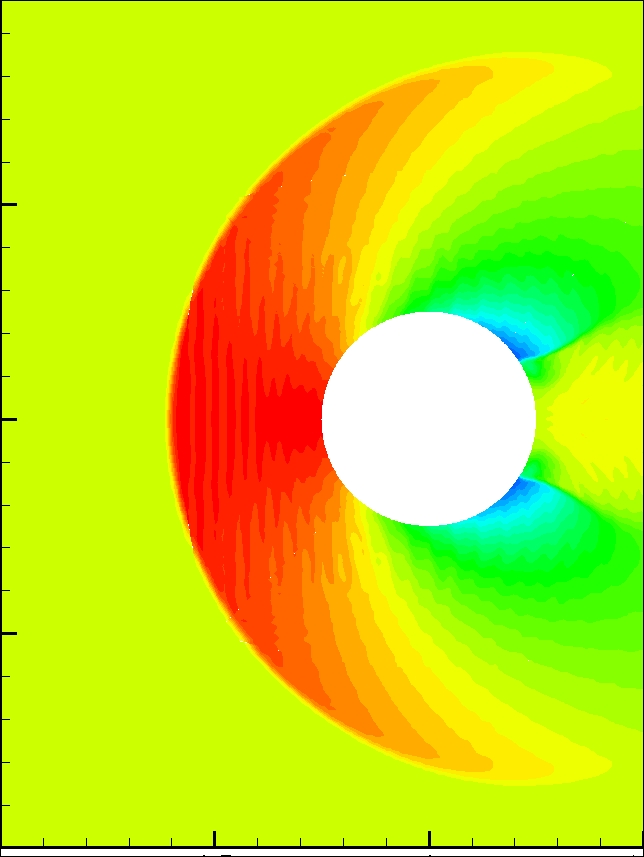}
\caption{Resulting density for transonic flow through circular cylinder using proposed PINN-WE (left) and WENO-Z method (right).} 
\end{figure}

 \begin{figure}
\includegraphics[width = 9cm]{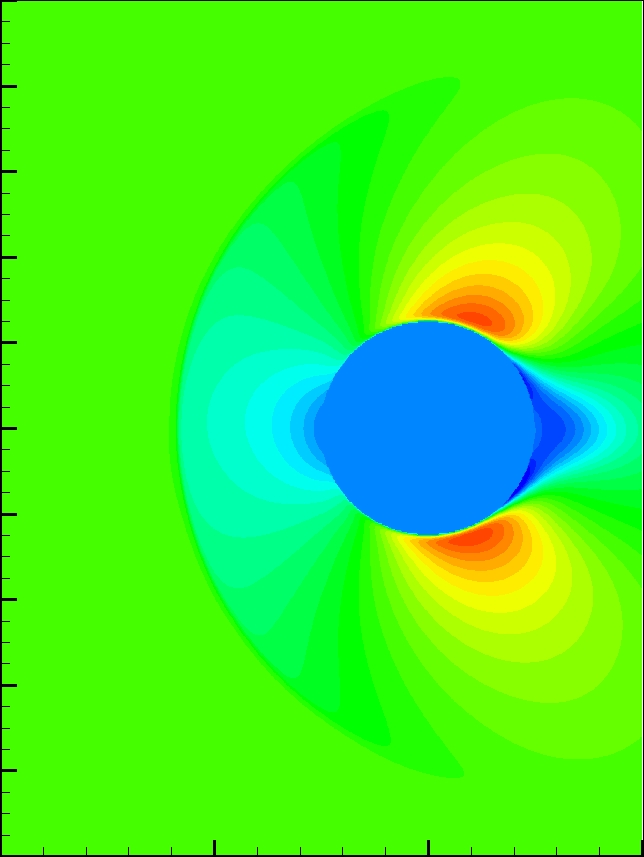}
\includegraphics[width = 9cm]{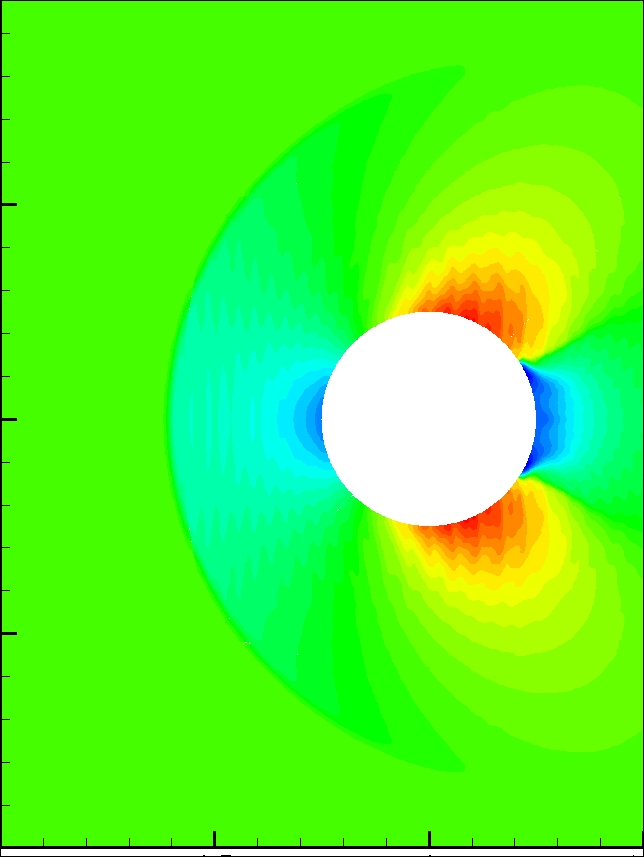}
\caption{Resulting velocity $u$ for transonic flow through circular cylinder using proposed PINN-WE (left) and WENO-Z method (right).} 
\end{figure}

 \begin{figure}
\includegraphics[width = 9cm]{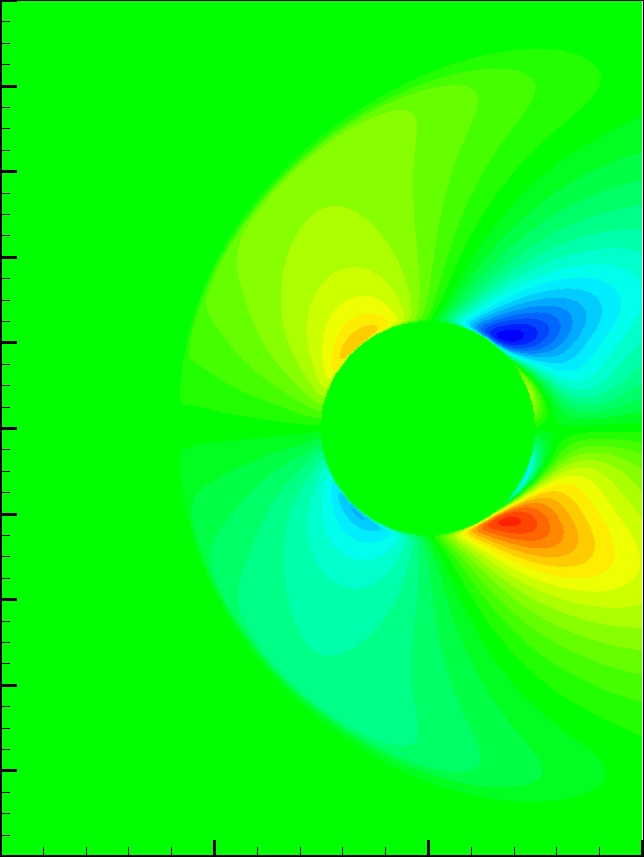}
\includegraphics[width = 9cm]{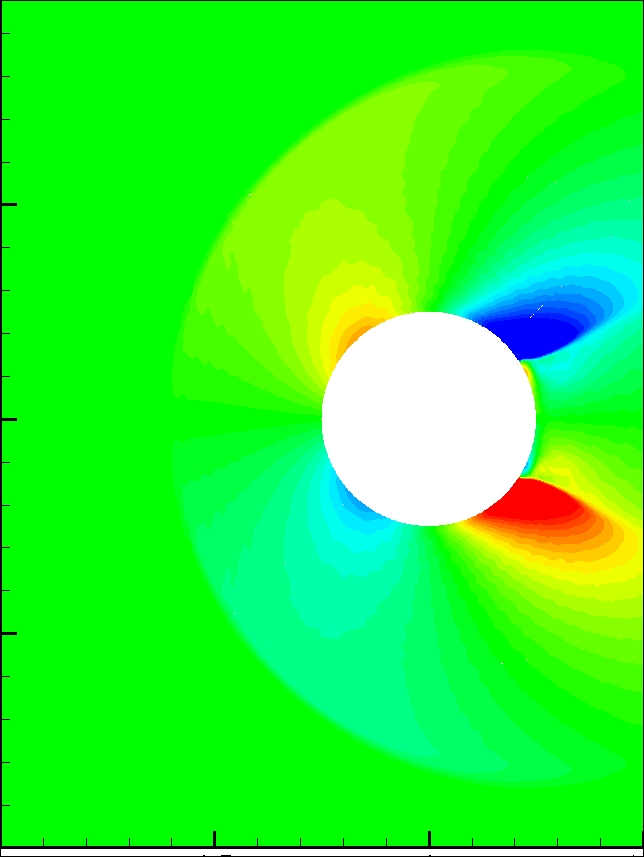}
\caption{Resulting velocity $v$ for transonic flow through circular cylinder using proposed PINN-WE (left) and WENO-Z method (right).} 
\end{figure}

 \begin{figure}
\includegraphics[width = 9cm]{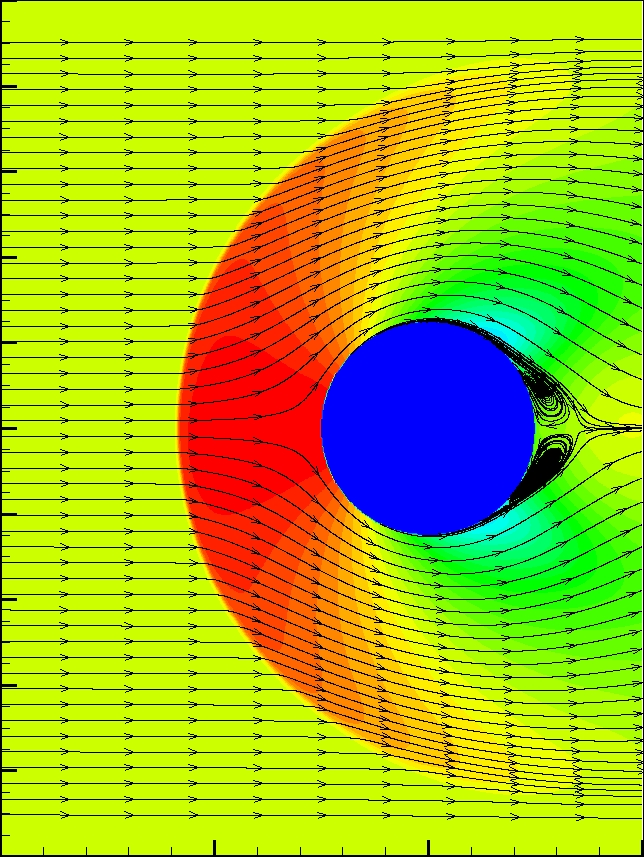}
\includegraphics[width = 9cm]{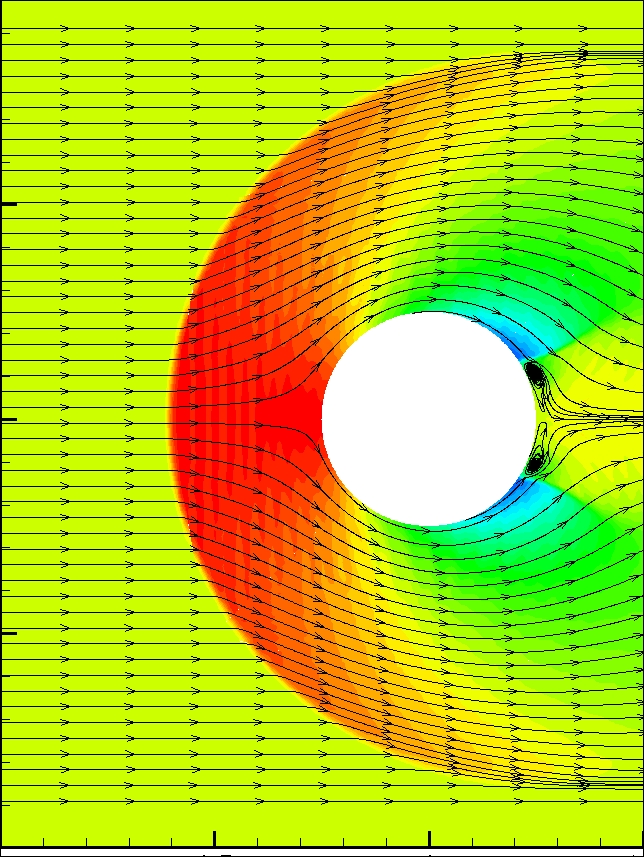}
\caption{Resulting streamline for transonic flow through circular cylinder using proposed PINN-WE (left) and WENO-Z method (right).} 
\end{figure}

 The results at time $0.4$ in Fig. 6 show that the proposed PINN-WE is also effective for relatively complex high-dimensional problems. Compared with the results from the high-order WENO-Z method with $100\times 100$ grids, more sharp shocks are obtained using our proposal. 

\section{Conclusions}
We propose a concept to capture discontinuities, especially shock waves, using a PINN for solving Euler equations. 
Unlike the idea of enhancing the NN expression in a large gradient domain, we consider the existence of a paradoxical problem within wrong inside shock points (transition-points). Regardless of increasing or decreasing gradients, these points likely increase the total loss. Thus, NN training may fall in conflict. Accordingly, we introduce a positive gradient-dependent weight into the governing equations to adjust the expression of a PINN in regions with different physical features. Then, for solving the Euler equations, we construct a weight inverse to the local physics compression by measuring the velocity divergence.

By solving the WEs using PINN, training focuses on smooth regions, while shock regions have very small weights. By relying on the real physics compression from the trained smooth regions, 
discontinuities automatically appear as the transition-points move out into smooth regions like passive particles.

We only focused on capturing discontinuities in this study, but many questions for solving Euler equations remain open. In fact, we found various problems that have not been addressed when simulating complex discontinuous problems. First, convergence to a real physical weak solution is difficult. Using WE, PINN can converge fast to a discontinuous result, but a nonphysical weak solution may be obtained for the Euler equations. Thus, effective entropy conditions and other physical limitations should be integrated into PINN. 
Second, the shock position may be inaccurate without the exact conservation of mass, momentum, and total energy in the PINN model. 
Third, as a global method, PINN can suitably simulate big structures but may inadequately reflect detailed structures (Fig. 5). Thus, the resolution of PINN should be improved to increase its applicability to diverse problems. 

\bibliography{mybibfile}
  \appendix

\end{document}